\documentclass[epj,numbook]{svjour}

\usepackage{amsmath,amssymb}
\usepackage{graphicx}
\usepackage{epsfig,rotate}
\usepackage{fancyheadings}

\begin{document}

\bibliographystyle{prsty}

\title{Exact results for the Kardar--Parisi--Zhang equation with spatially
  correlated noise}


\dedication{Dedicated to Franz Schwabl on the occasion of his 60th birthday.}

\author{H. K. Janssen\inst{1} \and U.C. T\"auber\inst{2} \and E. Frey\inst{2}}

\institute{Institut f\"ur Theoretische Physik III,
  Heinrich-Heine-Universit\"at, 40225 D\"usseldorf, Germany \\
  \email{janssen@thphy.uni-duesseldorf.de} \and Institut f\"ur Theoretische
  Physik T34,Technische Universit\"at M\"unchen,
  85747 Garching, Germany \\
  \email{taeuber@physik.tu-muenchen.de,frey@physik.tu-muenchen.de}}

\date{\today}

\abstract{\rm We investigate the Kardar--Parisi--Zhang (KPZ) equation in $d$
  spatial dimensions with Gaussian spatially long--range correlated noise ---
  characterized by its second moment $R(\vec{x}-\vec{x}') \propto 
  |\vec{x}-\vec{x}'|^{2\rho-d}$ --- by means of dynamic field theory and the
  renormalization group. Using a stochastic Cole--Hopf transformation we derive
  {\em exact} exponents and scaling functions for the roughening transition and
  the smooth phase above the lower critical dimension $d_c = 2 (1+\rho)$. Below
  the lower critical dimension, there is a line $\rho_*(d)$ marking the 
  stability boundary between the short-range and long-range noise fixed points.
  For $\rho \geq \rho_*(d)$, the general structure of the renormalization-group
  equations fixes the values of the dynamic and roughness exponents exactly, 
  whereas above $\rho_*(d)$, one has to rely on some perturbational techniques.
  We discuss the location of this stability boundary $\rho_* (d)$ in light of 
  the exact results derived in this paper, and from results known in the 
  literature. In particular, we conjecture that there might be two 
  qualitatively different strong-coupling phases above and below the lower 
  critical dimension, respectively.}


\titlerunning{Exact results for the KPZ equation with spatially
  correlated noise} \authorrunning{H.K. Janssen {\em et~al.}}

\maketitle

\section{Introduction}
\label{intro}

The dynamics of interfaces moving through random media has turned out to be one
of the most fascinating and at the same time challenging topics in theoretical
non-equilibrium physics. Among many other problems, kinetic roughening of
growing interfaces comprises one interesting class of such phenomena. Here the
randomness is described by an annealed random noise which mimics the random
adsorption of molecules onto a surface. The average force on the interface is
unimportant and may be removed from the equations of motion. Theoretical
modeling of those growth processes started with the work by Edwards and
Wilkinson (EW) \cite{edwards-wilkinson:82} who suggested that one might 
describe the dynamics of the height fluctuations by a simple linear diffusion
equation. Kardar, Parisi, and Zhang (KPZ) \cite{kardar-parisi-zhang:86}
realized that there is a relevant term proportional to the square of the height
gradient which represents a correction for lateral growth. What at first glance
seemed to be a simple generalization of the diffusion equation became an
archetypical example of non-equilibrium dynamic phenomena
\cite{meakin:93,halpin-healy-zhang:95,krug:97}.

Despite its fundamental importance as a generic theoretical model for
non-equilibrium phase transitions, to date there seems to be no experimental
system which quantitatively confirms the predictions of the KPZ model 
satisfactorily \cite{halpin-healy-zhang:95,krug:97}. This discrepancy has
inspired the introduction of various modifications of the original model. There
are several ways in which the KPZ equation provides an incomplete description
of an actual experimental system: First of all, one might question the validity
of assuming uncorrelated Gaussian noise. In a real system, the noise could be 
correlated \cite{medina-hwa-kardar-zhang:89}, non-Gaussian \cite{zhang:90}, or 
even quenched \cite{csahok-honda-vicsek:93,amaral-barabasi-stanley:94} instead 
of annealed. Next, in many physical processes there exist long-range 
hydrodynamic interactions leading to non-local dynamics for the height 
fluctuations \cite{krug-meakin:91,mukherji-bhattacharjee:97}. Finally, there 
may be other non-linearities which describe important physics omitted in the 
KPZ equation, in particular in the strong-coupling rough phase.

Quite generally, in a genuine non-equilibrium system, the form of the noise
correlations in an effective Langevin-type description is a crucial ingredient
of the modeling. This is in contrast to equilibrium dynamics, where the 
functional form of the noise as well as its strength are fixed via an Einstein
relation (detailed-balance condition) which ensures that asymptotically the
probability distribution will be the equilibrium one. Uncorrelated white noise
is often a straightforward choice, but the sensitivity and stability of the
ensuing results need to be carefully tested against modifications of the noise
correlations. In terms of the renormalization-group (RG) language, one must
establish which modifications of the noise correlators constitute relevant
perturbations, and may thus lead to novel universality classes.

In this work, we focus on the effect of {\em spatially correlated Gaussian
noise} $\tilde \zeta (\vec{x},t)$ with
\begin{equation}
  \langle {\tilde \zeta}(\vec{x},t) {\tilde \zeta}(\vec{x}',t') \rangle 
  = 2 R (\vec{x}-\vec{x}') \, \delta(t-t') \ ,
\end{equation}
where for some part of the paper we will restrict ourselves to the particular 
form  
\begin{equation}
  R ( \vec{x}-\vec{x}' ) \propto  |\vec{x}-\vec{x}'|^{2\rho-d} 
\label{noidef}
\end{equation}
with $d$ being the spatial dimension of the surface. Our motivation, however, 
is not to introduce a realistic growth model which allows for a quantitative
comparison with experiments, but to gain a better understanding of the KPZ
equation and possibly reveal some of its hidden secrets. Since noise
correlations are actually generated by the non-linearity in the KPZ equation,
one might hope that via introducing such correlations to begin with, one could
learn something about the notorious strong-coupling behavior of the rough 
phase, separated from the smooth EW regime through a second-order dynamic
roughening transition (above the lower critical dimension $d_c$).

The KPZ equation with correlated Gaussian noise has first been studied by
Medina {\em et\/ al.} \cite{medina-hwa-kardar-zhang:89}, applying dynamic
renormalization-group (RG) methods. Among other results, they found in a 
one-loop perturbative renormaliza\-tion-group analysis that for the 
($1+1$)-dimensional growth model, there exists a critical value of the 
correlation exponent, $\rho_c = \frac{1}{4}$, below which long-range noise 
correlations are {\em irrelevant} and the roughness and dynamic critical 
exponents retain their values $\chi = \frac{1}{2}$ and $z=\frac{3}{2}$ 
associated with uncorrelated (white) noise. Above $\rho_c$, long-range noise 
correlations become {\em relevant} in the RG sense, and a novel fixed point 
emerges, at which the exponents assume the values 
\begin{equation}
  \chi = \frac{1+2 \rho}{3} \, , \quad \text{and} \quad 
  z = \frac{5-2 \rho}{3} \, .
\label{1D:1-loop-exponents}
\end{equation}
These results have actually been found already by Kardar
\cite{kardar:87} and Nattermann \cite{nattermann:87} in the context of
a generalized interface model for domain wall roughening which
interpolates between a random bond ($\rho = 0$) and random field
($\rho = 1$) Ising model; some limiting results were also obtained by
Zhang \cite{zhang:86} building on
Ref.~\cite{forster-nelson-stephen:77}. Eq.~\ref{1D:1-loop-exponents}
also corresponds to an estimate for the exponents obtained by Flory
scaling \cite{kardar:87}.  In addition, these results are in accord
with a functional renormaliza\-tion-group calculation by Halpin-Healy
\cite{halpin-healy:89,halpin-healy:90}.  The above values are,
however, in conflict with results obtained by means of a replica
scaling method \cite{zhang:90b}, which give critical exponents quite
different from (\ref{1D:1-loop-exponents}), and would predict a
critical value of $\rho_c = \frac{1}{2}$.

To date, there have also been several numerical investigations to determine the
dependence of the critical exponents on the correlation exponent of the noise
$\rho$. For $d=1$, early simulations \cite{amar-lam-family:91} using restricted
solid-on-solid (RSOS) and ballistic deposition models, seemed to strong\-ly
support the predictions of the dynamic RG calculation with $\rho_c = 
\frac{1}{4}$. These results, however, could not be confirmed by later 
simulations using complementary models \cite{peng-havlin-schwartz-stanley:91}.
Employing an alternative scheme to generate correlated noise, Pang 
{\em et\/ al.} \cite{pang-yu-halpin_healy:95} carried out extensive numerical 
studies on the RSOS model and the presumably equivalent model of a directed 
polymer in a random medium (DPRM). It seems that the exponents obtained from 
the simulation of the RSOS model deviate from the results for the short-range 
fixed point already at $\rho \approx \frac{1}{8}$; the DPRM data break off even
earlier. In summary, there is no coherent picture emerging from the numerical 
simulations performed on the basis of an interface-height or directed-polymer 
representation, respectively. We suppose that these discrepancies between
different simulations may partly be traced back to the presence of 
corrections-to-scaling effects due to the crossover between the short-range and
long-range fixed point. Our impression is that a careful re-analysis of the 
numerical simulations in the light of the results presented in this paper would
be highly desirable. 

Hayot and Jayaprakash \cite{hayot-jayaprakash:96,hayot-jayaprakash:97} have 
recently investigated the effect of correlated noise in the framework of the 
{\em stoch\-astic Burgers equation}. They find very clear evidence that the 
critical value for $\rho$ is $\rho_c = \frac{1}{4}$, and that for $\frac{1}{4} 
\leq \rho \leq 1$ the asymptotic behavior is within numerical errors described 
by the exponents derived from the one-loop RG calculation, 
(\ref{1D:1-loop-exponents}). The most surprising and remarkable result of this 
paper is that the numerical simulations show quantitative agreement with the 
one-loop results for the roughness exponent $\chi$ even for $\rho \geq 1$, 
i.e., in a parameter regime which (because of higher-order non-linear terms) 
lies outside the realm of the RG analysis. At these large values of $\rho > 1$,
the velocity fields $u(x,t)$ in the statistically stationary state display 
well-defined shocks leading to multifractality and anomalous dynamics. The 
truncated velocity correlation function $\langle [u(x,t)-u(x,0)]^2 \rangle$ 
displays two distinct scaling regimes. At short times, the dynamic exponent $z$
is given by the one-loop result (\ref{1D:1-loop-exponents}), while 
$z \approx 1$ at longer times. The value $z=1$ is a consequence of the presence
and ballistic motion of the shock fronts.  Through an analysis of the scaling 
behavior of appropriate composite operators, it is found that these 
higher-order operators do not become increasingly relevant. This suggests the 
validity of simple balancing arguments \cite{hayot-jayaprakash:96}, which lead 
to exactly the same exponent relations (\ref{1D:1-loop-exponents}) as the 
one-loop result obtained by Medina {\em et\/ al.} 
\cite{medina-hwa-kardar-zhang:89}. 

Furthermore, the $(2+1)$-dimensional KPZ model was investigated by direct 
numerical integration of the KPZ equation \cite{li:97}. Similar to the case 
$d=1$, there seems to be a critical value $\rho_c \approx 0.5$ below which the
exponents for uncorrelated noise are oberved, and above which a strong 
$\rho$-dependence of the critical exponents is found. Since a one-loop 
approximation \cite{medina-hwa-kardar-zhang:89} does not lead to a stable fixed
point in two dimensions, a comparison with analytic results was not possible up
to now. We will see later in this paper that our analysis not only allows for 
such a comparison, but even leads to {\em exact} expressions for the critical
exponents (provided the associated finite RG fixed points exist).

The outline of the paper is as follows. In the subsequent section, we set up 
the field-theoretical tools (such as the dynamic functional, renormalization
factors, renormaliza\-tion-group equations, and scaling relations) for the
analysis of the KPZ equation with correlated noise. This section contains
mostly definitions and recapitulations, as well as discussions of results known
in the literature \cite{medina-hwa-kardar-zhang:89} in light of the 
field-theory framework. In particular, we present a scrutinized discussion of 
the derivation of the dynamic functional from the stochastic equation of 
motion, which pays attention to some subtleties often neglected in the 
literature, yet of considerable importance for the stochastic Cole-Hopf
transformation to be introduced in section \ref{sec:cole_hopf}. Section 
\ref{1dren} then deals with an analysis of the critical behavior below the 
lower critical dimension $d_c = 2 (1 + \rho)$ where there is only a rough 
phase, and no roughening transition from a smooth into a rough phase exists. 
Since within a formulation in terms of the height variables $h(\vec{x},t)$, 
there is no critical dimension around which one could try to perform an 
$\varepsilon$ expansion, we have to rely on a fixed-dimension loop expansion 
which necessarily becomes non-minimal in the sense that UV-finite terms in the
renormalization constants need to be retained. Notice, however, that there is 
one special point in the $(\rho,d)$--plane, corresponding to the Burgers 
equation with non-onserved noise, which actually allows for an 
$\varepsilon$ expansion around the upper critical dimension $4$ 
\cite{forster-nelson-stephen:77}. Here we show how one may perform a 
fixed-dimension expansion below $d_c$, the applicability of which, however,
seems mainly be restricted to the $(1+1)$-dimensional situation, where the 
critical behavior can also be analyzed by starting from a driven diffusion 
equation. 

The behavior of the roughening transition above the critical dimension $d_c$ is
studied in section \ref{sec:rough} by means of a stochastic Cole--Hopf 
transformation, which transforms the dynamic functional of the KPZ equation
into a dynamic functional whose structure is very reminiscent of the field
theory for diffusion-limited pair annihilation 
\cite{peliti:84,lee:94,lee-cardy:95}. This allows us to present {\em exact} 
expressions for the entire set of renormalization-group flow functions, in
particular the RG beta functions. We find that the roughening transition in the
presence of correlated noise is characterized by the dynamic exponent 
$z_c = 2$, the roughness exponent $\chi_c = 0$, the crossover exponent 
$\phi_c = d-2$, and the correction-to-scaling exponent 
$\omega_c = d - 2(1+\rho)$.  Above the lower critical dimension $d_c$, the 
long-range scaling fixed point is stable, and the smooth phase is described by 
the scaling exponents $z_{\rm sm} = 2$, $\chi_{\rm sm} = 1+\rho-\frac{d}{2}$, 
and the {\em different} correction-to-scaling exponents 
$\omega_1 = d-2(1+\rho)$, $\omega_2 = 2(1+2\rho)-d$ for $2(1+\rho) < d < 
2(1+2\rho)$, and $\omega_1 = d-2(1+2\rho)$, $\omega_2 = 2\rho$ for $d \geq 
2(1+2\rho)$, respectively. The rough phase is found to be non-accessible 
through perturbational means. Finally, for $d > d_c$ the theory can be 
generalized to an arbitrary noise correlator (albeit local in time). The 
ensuing Bethe--Salpeter integral equation is analyzed numerically for Gaussian 
and power-law noise spectrum in section \ref{sec:bseqn}. We conclude with a 
brief summary, and a discussion of some of the open problems.

\section{The KPZ field theory with spatially correlated noise}
\label{gener}

\subsection{Model equations and field theory}
\label{model_equations}

In the standard convention, the Kardar--Parisi--Zhang or KPZ equation for the
interface height fluctuation $h(\vec{x},t)$ in $d+1$ dimensions is written in
the form \cite{kardar-parisi-zhang:86,medina-hwa-kardar-zhang:89}
\begin{equation}
  \frac{\partial h}{\partial t} = D \vec{\nabla}^2 h + 
  \frac{\lambda}{2} (\vec{\nabla} h)^2 + {\tilde \zeta} \ ,
\label{eq:kpz_standard}
\end{equation}
where $D$ denotes the effective interface surface tension, or, in the Burgers
representation, the kinematic viscosity of the potential velocity field 
${\vec u} \propto \vec{\nabla} h$. The coefficient $\lambda$ describes the 
strength of the non-linearity, associated with either the curvature-driven 
growth process or the convective term in the hydrodynamic equation, 
respectively.

Furthermore, $\tilde \zeta$ denotes a stochastic driving force with zero mean,
$\langle \tilde \zeta (\vec{x},t ) \rangle$, and Gaussian correlations
determined by the second moment
\begin{equation}
  \langle \tilde{\zeta} \left( \vec{x},t \right) \tilde{\zeta} \left( 
  \vec{x}^\prime ,t^\prime \right) \rangle = 2 \widetilde{R} \left( 
  \vec{x}-\vec{x}^\prime \right) \delta \left( t-t^\prime \right) \ , \label{1}
\end{equation}
which we assume to be local in time but long-range correlated in space. In
particular, we are going to consider the case where in Fourier representation
\begin{equation}
  \widetilde{R} \left( \vec{x} \right) = \int \frac{d^q q}{(2 \pi)^d} \, 
  \widetilde{R} \left( \vec{q}\right) \exp \left( i{\bf q\cdot x} \right)
\end{equation}
takes the form
\begin{equation}
  \widetilde{R} \left( \vec{q} \right) = \widetilde{D} 
  \left( 1+wq^{-2\rho } \right) \ .
\label{2}
\end{equation}
Here, in addition to the short-range term proportional to $\widetilde{D}$, we 
have allowed for a power-law long-range contribution to $\widetilde{R} \left(
\vec{x} \right) \propto x^{2\rho -d}$ with amplitude $\widetilde{\Delta} =
\widetilde{D} w$ \cite{medina-hwa-kardar-zhang:89}.

For the field-theoretic analysis, and a direct transfer of standard results
from equilibrium critical dynamics to the analysis of the above non-equilibrium
Langevin equation, it is more appropriate to rescale the fluctuating and noise 
fields according to $s(\vec{x},t) = (D/{\widetilde D})^{\frac{1}{2}} h(\vec{x},
t)$ and $\zeta(\vec{x},t) = (D/{\widetilde D})^{\frac{1}{2}} {\tilde \zeta}
(\vec{x},t)$, which leads to
\begin{equation}
  \frac{\partial s}{\partial t} = D \vec{\nabla}^2 s + 
  \frac{D g}{2} (\vec{\nabla} s)^2 + \zeta \ , 
\label{kpzeqn}
\end{equation}
where $g = \lambda ({\widetilde D} / D^3)^{\frac{1}{2}}$ represents the
effective non-linear coupling of the KPZ problem. In this reparametrisation, 
the noise correlations in Fourier space read
\begin{eqnarray}
  &&\langle \zeta(\vec{q},\omega) \zeta(\vec{q}',\omega') \rangle = 
  2 D \left( 1 + w q^{-2\rho} \right) \nonumber \\
  &&\qquad\qquad\qquad\quad \times (2\pi)^d \delta(\vec{q}+\vec{q}')
  \delta(\omega+\omega') \ ,
\label{kpzcrq}
\end{eqnarray}
with $w$ characterizing the relative strength of the spatially long-range and
local noise correlations, respectively. Notice that through this simple 
rescaling, we have rendered the local-noise amplitude identical with the
kinetic coefficient in the Langevin equation. However, this does {\em not} in
general imply, even for $w = 0$, that detailed balance holds. For, despite the
validity of the Einstein relation for $w = 0$, the integrability condition
\cite{deker-haake:75} for the reversible force term $D g (\vec{\nabla} S)^2/2$ 
in Eq.~(\ref{kpzeqn}) is fulfilled in one dimension {\em only}. It is well
known that in this case the stationary probability distribution is given by
\begin{equation}
  {\cal P}_{st}\left[ s\right] \propto 
  \exp \left( -{\cal H}\left[ s\right] \right) =
  \exp \left( -\frac{1}{2}\int_x \left( \vec{\nabla} s\right) ^{2}\right) \, ,
\label{eq:1dstat}
\end{equation}
where we have introduced the short-hand notation $\int_x = \int d^dx$.

Langevin equations like the KPZ equation (\ref{kpzeqn}) are not well-suited for
the development of a systematic renormalized field theory. Furthermore, their
interpretation depends on the underlying time discretization. Here, we shall 
use an interpretation in the sense of Ito. Instead of the Langevin equation, we
write down the corresponding Gaussian transition probability distribution 
${\cal P}\left[ s\left( t+\tau \right) |s\left( t\right) \right]$ from a
configuration $\left\{ s\left( \vec{x},t\right) \right\} $ at time $t$ to a
configuration $\left\{ s\left( \vec{x},t+\tau \right) \right\} $ at time 
$t+\tau$, where $\tau $ is a small time increment,
\begin{eqnarray}
  &&{\cal P}\left[ s\left( t+\tau \right) |s\left( t\right) \right] \propto 
  \nonumber \\
  &&\exp \Bigl( -\frac{1}{4\tau} \int_{x, x^{\prime }}
  \bigl[ s\left( \vec{x},t+\tau \right) - s \left(\vec{x},t\right) - 
  \tau V\left(s\left(\vec{x},t\right) \right) \bigr] \nonumber \\
  &&\times R^{-1}\left( \vec{x}-\vec{x}^{\prime }\right) 
  \bigl[ s\left( \vec{x}^{\prime },t+\tau \right) 
  -s\left( \vec{x}^{\prime },t\right) 
  -\tau V\left(s\left( \vec{x}^{\prime },t\right) \right) \bigr] \Bigr)
  \nonumber \\
  &&= \int {\cal D} \left[ \tilde{s}\left( t+\tau \right) \right] \,
  \exp \left( -\int_x \tilde{s}\left( \vec{x},t+\tau \right) 
  \bigl[ s\left( \vec{x},t+\tau \right) \right. \nonumber \\
  &&\qquad \qquad \qquad \qquad \qquad
  -s\left( \vec{x},t\right) -\tau V\left( s\left( \vec{x},t\right) \right) 
  \bigr] \nonumber \\
  &&\left. +\tau \int_{x, x^{\prime}} \tilde{s}\left( \vec{x},t+\tau \right) 
  R\left( \vec{x}-\vec{x}^{\prime }\right) \tilde{s}
  \left( \vec{x}^{\prime },t+\tau \right) \right) \ .  
\label{eq:prob_distribution}
\end{eqnarray}
Here we have defined 
\begin{eqnarray}
  V\left( s\right) =D\nabla ^{2}s+\frac{Dg}{2}\left( \vec{\nabla} s\right) ^{2}
\end{eqnarray}
and with $R^{-1} (\vec{x})$ denote the reciprocal kernel of $R$, i.e., $\int
d^{d}x^{\prime \prime }\,R\left( \vec{x}-\vec{x}^{\prime \prime }\right)
\,R^{-1}\left( \vec{x}^{\prime \prime }-\vec{x}^{\prime }\right) =\delta \left(
\vec{x}-\vec{x}^{\prime }\right) $. We have al\-so introduced the purely 
imaginary response field $\tilde{s}\left( \vec{x},t\right) $ which corresponds
to the adjungated operator in the operator formulation of statistical dynamics
by Martin, Siggia, and Rose \cite{martin-siggia-rose:73}. Note that the last
expression in (\ref{eq:prob_distribution}) is properly normalized,
\begin{equation}
\int {\cal D}\left[ s\left( t+\tau \right) \right] {\cal P}\left[ s\left(
t+\tau \right) |s\left( t\right) \right] =1 \ ,
\label{6}
\end{equation}
if one includes a purely numerical constant, independent of any parameter of 
the theory, in the functional measure ${\cal D}\left[ \tilde{s}\left( t\right),
s\left( t\right) \right] $. The joint probability distribution can be expressed
in terms of a Markoff chain 
\begin{eqnarray}
  \prod_{k=1}^{K}{\cal P}\left[ s\left(
    t+k\tau \right) |s\left( t+\left( k-1\right) \tau \right) \right] \ .
\label{eq:markov_chain}
\end{eqnarray}
Upon taking the limit $\tau \rightarrow 0$, $K \rightarrow \infty $ with $\tau
K = T ={\rm const.}$, we finally arrive at the path probability distribution
\begin{eqnarray}
  &&{\cal P}\left[ s,0<t\leq T|s\left( t=0\right) = s_0 \right] \nonumber \\
  &&\quad = \left. \int 
  {\cal D}\left[ \tilde{s}\right] \,\exp \left( -{\cal J}_{0}^{T}\left[ 
  \tilde{s},s\right] \right) \right| _{s\left( t=0\right) =s_{0}}  \label{7}
\end{eqnarray}
for the field configurations $\left\{ s\left( \vec{x},t\right) \right\} $ in
the time interval $0<t\leq T$, starting from a given initial configuration 
$\left\{ s\left( \vec{x},0\right) =s_{0}\left( \vec{x}\right) \right\} $. The 
dynamic functional \cite{janssen:76,dominicis:76,janssen:79,janssen:92} is 
given by
\begin{eqnarray}
  {\cal J}_{0}^{T}\left[ \tilde{s},s\right] 
  &=&\int_{0}^{T} dt \int d^{d}x
   \biggl\{  \tilde{s}\left( \vec{x},t\right) 
   \Bigl[ \dot{s}\left( \vec{x},t\right) 
   -D\vec{\nabla}^{2}s\left( \vec{x},t\right) \nonumber \\
  &&\qquad \qquad \qquad \qquad \quad
  -\frac{Dg}{2}\left( \vec{\nabla} s\left( \vec{x},t\right) \right)^{2} \Bigr] 
  \nonumber \\
  &&- \int d^{d}x^{\prime }\,\tilde{s}\left( \vec{x},t\right)
  R\left( \vec{x}-\vec{x}^{\prime }\right) 
  \tilde{s}\left( \vec{x}^{\prime},t\right) \biggr\} \ .  
\label{eq:dyn_functional}
\end{eqnarray}
Expectation values can now be calculated by means of path integrals with the 
exponential weight $\exp \left( -{\cal J}\left[ \tilde{s},s\right] \right)$.
Perturbational expansions are then to be generated in usual diagrammatical 
terms, and form the basis for a renormalized dynamical field theory
\cite{bausch-janssen-wagner:76,dominicis-peliti:78}.

But only in $d=1$ and without correlated noise can the dynamic functional be
written in the detailed-balance form 
\begin{equation}
  {\cal J}_{0}^{T}\left[ \tilde{s},s\right] =\int_{0}^{T}dt\int dx\,\left\{ 
  \tilde{s}\left[ \dot{s}+{\cal M}\frac{\delta {\cal H}}{\delta s}\right] -
  \tilde{s}{\cal M}\tilde{s}\right\} \ ,  \label{10}
\end{equation}
where the integral-kernel ${\cal M}$ is defined by 
\begin{eqnarray}
  {\cal M}\left( x,y\right) 
  &=& D\left[ \delta \left( x-y\right) +\frac{g}{3}
  \left( \theta \left( x-y\right) \frac{\partial s\left( x\right) }{\partial x}
  \right. \right. \nonumber \\
  &&\left. \left.
  -\theta \left( y-x\right) \frac{\partial s\left( y\right) }{\partial y}
  \right) \right] \ ,  \label{11}
\end{eqnarray}
with the hamiltonian ${\cal H}$ given by Eq.~(\ref{eq:1dstat}). Obviously, 
${\cal J}$ now obeys the time inversion symmetry \cite{janssen:79,janssen:92}
\begin{eqnarray}
  s\left( x,t\right) &\rightarrow & s\left( x,-t\right) , \nonumber \\
  \tilde{s}\left( x,t\right) &\rightarrow &-\tilde{s}\left( x,-t\right)
  +\left. \frac{\delta {\cal H}}{\delta s}\right| _{x,-t} \nonumber \\
  &&=-\tilde{s}\left(x,-t\right) - \vec{\nabla}^{2}s\left( x,t\right) \ ,
  \nonumber \\
  {\cal J}_{0}^{T}+\left. {\cal H}\right| _{t=0} 
  &\rightarrow &{\cal J}_{-T}^{0}+\left. {\cal H}\right| _{t=-T} \ . \label{12}
\end{eqnarray}
As the transition probability distribution is given by the path integral
\begin{eqnarray}
  {\cal P}\left[ s_{T},T|s_{0},0\right] 
  &=&\int {\cal D}\left[ \tilde{s},s\right] 
  \,\prod_{x}\delta \bigl( s\left( x,T\right) =s_{T}\left( x\right) \bigr)
  \nonumber \\
  &&\left. \!\! \times 
  \exp \left( -{\cal J}_{0}^{T}\left[ \tilde{s},s\right] \right)
  \right|_{s\left( x,0\right) = s_{0}\left( x\right) } \; , \label{13}
\end{eqnarray}
one derives directly from (\ref{12}) the detailed balance condition
\begin{equation}
  {\cal P}\left[ s_{T},T|s_{0},0\right] {\cal P}_{st}\left[ s_{0}\right] =
  {\cal P}\left[ s_{0},0|s_{T},-T\right] {\cal P}_{st}\left[ s_{T}\right] \ ,
\label{14}
\end{equation}
which implies that ${\cal P}_{st}\left[ s\right]$ in (\ref{eq:1dstat}) is
indeed the stationary probability distribution for $d=1$. 
In Appendix~\ref{appa}, we present an alternative derivation of this result on
the basis of the associated Fokker--Planck equation.

In the following we shall not be interested in the initial-time properties of 
the stochastic process. Thus we extend the time interval in the expression for 
the dynamic functional from minus to plus infinity, which then reduces to
\begin{eqnarray}
  &&{\cal J} = \int_x \int_t {\tilde s} \left[ \partial_t s
    - D \vec{\nabla}^2 s
    - \frac{D g}{2} (\vec{\nabla} s)^2 \right] \nonumber \\ 
  &&- \int_q  \int_\omega  D \, {\tilde s} (-\vec{q},-\omega) 
  \left( 1 + w q^{-2\rho} \right) {\tilde s} (\vec{q},\omega) \ .
\label{dynfun}
\end{eqnarray}
Here, we have employed the short-hand notation $\int_q \ldots = \int \ldots
d^dq / (2\pi)^d$ and $\int_\omega \ldots = \int \ldots d\omega / 2\pi$.

We finally note that the dynamic functional is invariant under Galilean 
transformations
\begin{eqnarray}
  s\left( \vec{x},t\right) &\rightarrow &s\left( \vec{x}+Dg\vec{v}t,t\right) +
  {\bf v\cdot x} \ ,  \nonumber \\
  \tilde{s}\left( \vec{x},t\right) &\rightarrow &\tilde{s}\left( \vec{x}+Dg
  \vec{v}t,t\right) \ .  
\label{eq:galilean}
\end{eqnarray}
Implications of this Galilean invariance, specifically on the renormalizability
of the KPZ dynamic field theory, are discussed in Appendix~\ref{appb}.

\subsection{Renormalization and scaling}

We proceed by introducing renormalized quantities and the associated
renormalization constants. In the ``standard'' representation of the KPZ
problem (\ref{eq:kpz_standard}), there appears no field renormalization 
\cite{frey-taeuber:94}. Furthermore, Galilean invariance of the noisy Burgers 
equation, or infinitesimal interface tilt invariance for the KPZ problem, 
implies that the non-linear coupling $\lambda$ does not renormalize either, 
which leaves only the {\em two} independent renormalization factors for the 
diffusion constant $D$ and the noise strength ${\widetilde D}$ to be 
determined. This remains true even for the situation with power-law correlated
spatial noise, for which Galilean invariance of course still holds. 
Furthermore, the long-range noise contributions do enter the renormalization of
the diffusion constant and the short-range noise amplitude (see 
Sec.~\ref{1dren}); in fact, even if ${\widetilde D}$ were zero initially, such
spatially local noise terms would be generated under renormalization.  However,
the amplitude of the spatially correlated noise ${\widetilde \Delta}$ itself is
{\em not} renormalized, because non-analytic terms of the form $\propto 
q^{-2\rho}$ can never be produced in a perturbation expansion for the 
corresponding vertex function (evaluated at non-zero external wave vector $q$).

In our representation in section \ref{model_equations}, on the other hand, we
have absorbed the ratio $D / {\widetilde D}$ in the new fields $s$, which leads
to a non-trivial field renormalization \cite{fnote2}. However, the diffusive
character of the fluctuations, and the wave-vector dependence of the non-linear
vertices still implies the {\em exact} result (to all orders in perturbation 
theory) \cite{frey-taeuber:94,taeuber-frey:95} 
\begin{equation}
  \Gamma_{{\tilde s}s}(\vec{q}=\vec{0},\omega) \equiv i \omega \ ,
\label{q0verf}
\end{equation}
and we can therefore define renormalized fields according to
\begin{eqnarray}
  s &\rightarrow& s_0 = Z^{\frac{1}{2}} \, s \ , \nonumber \\
   {\tilde s} &\rightarrow& {\tilde s}_0 = Z^{-\frac{1}{2}} \, {\tilde s} \ .
\label{fldren}
\end{eqnarray}
The above considerations still remain valid, and we merely need to introduce
one more independent renormalization constant $Z_D$ via
\begin{equation}
  D \rightarrow D_0 = Z \, Z_D \, D \ .
\label{difren}
\end{equation}
The non-renormalization of the three-point vertex and the non-analytic 
long-range noise amplitude then implies that we can introduce corresponding
dimensionless renormalized parameters $u$ and $v$ via
\begin{equation}
  g^2 \rightarrow g_0^2 = Z^{-3} \, Z_D^{-2} \, u \, A_d^{-1} \, \mu^{2-d} \ ,
\label{verren}
\end{equation}
and
\begin{equation}
  w g^2 \rightarrow 
  w_0 \, g_0^2 = (Z \, Z_D)^{-3} \, v \, B_d^{-1} (\rho) \, 
  \mu^{2(1+\rho)-d} \ .
\label{noiren}
\end{equation}
Eqs.~(\ref{verren}) and (\ref{noiren}) define the two effective couplings
associated with the combination of the non-linearity in the Langevin equation
with the local and power-law noise correlations, respectively. For convenience,
we shall choose the following geometric factors 
\begin{equation}
  A_d := \frac{\Gamma(2-\frac{d}{2})}{2^{d-1} \pi^{\frac{d}{2}} d} \, 
         \sin \left( \frac{\pi d}{4} \right)\ ,
\label{geofcA}
\end{equation}
and
\begin{equation}
    B_d (\rho) := \frac{\Gamma(2+\rho-\frac{d}{2}) \Gamma(\frac{d}{2}-\rho)}
    {2^{d-1} \pi^{\frac{d}{2}} d \, \Gamma(\frac{d}{2})} \
    \sin \left( \frac{\pi (d - 2 \rho)}{4} \right) \ ,
\label{geofcB}
\end{equation}
where $\Gamma(x) = \Gamma(1+x) / x$ denotes Euler's gamma function. The
momentum scale $\mu$ represents an appropriately chosen renormalization point
outside the infrared-singular region, e.g., $\vec{q} = \vec{0}$, and 
$\omega/2D = \mu^2$.

The asymptotic scaling behavior of the theory is governed by the properties in
the vicinity of the renormaliza\-tion-group fixed points $u_*$ and $v_*$,
defined as the zeros of the RG beta functions
\begin{equation}
  \beta_u = \mu \, \frac{\partial u}{\partial \mu} \bigg\vert_0 \ , 
  \quad {\rm and} \quad 
  \beta_v = \mu \, \frac{\partial v}{\partial \mu} \bigg\vert_0 \ ,
\label{betfun}
\end{equation}
which describe the dependence of the renormalized non-linear couplings on the
momentum scale $\mu$; the above derivatives are to be taken after re-expressing
$u$ and $v$ in terms of their bare counterparts $g_0^2$ and $w_0 g_0^2$. In 
addition, we define the Wilson functions
\begin{equation}
  \gamma = \frac{\partial \ln Z}{\partial \ln \mu} \bigg\vert_0 \ ,
  \quad  \quad
  \gamma_D = \frac{\partial \ln Z_D}{\partial \ln \mu} \bigg\vert_0 \ ,
\label{zetfun}
\end{equation}
yielding the anomalous dimensions for the field $s$ and the diffusion constant
$D$, respectively, and
\begin{equation}
  \zeta =  \frac{\partial \ln D}{\partial \ln \mu} \bigg\vert_0 
        = - \gamma - \gamma_D \ .
\end{equation}
Eqs.~(\ref{verren}) and (\ref{noiren}) then imply the following relations
\begin{eqnarray}
  \beta_u &=& \left( d - 2 + 3 \, \gamma + 2 \, \gamma_D \right) u \ , 
\label{betag2} \\
  \beta_v &=& \left( d - 2 - 2 \rho + 3 \, \gamma + 3 \, \gamma_D \right) v \ .
\label{betaf2}
\end{eqnarray}

The fact that the {\em bare} correlation function $C_0(x,t)$ is of course
independent of the chosen normalization point $\mu$, leads to the 
Gell-Mann--Low renormalization-group equation (see, e.g., 
Refs.~\cite{janssen:76,frey-taeuber:94} )
\begin{eqnarray}
  &&\left[ \mu \, \frac{\partial}{\partial \mu}
  + \gamma_D \, D \, \frac{\partial}{\partial D} 
  + \beta_u \, \frac{\partial}{\partial u} 
  + \beta_v \, \frac{\partial}{\partial v} + \gamma \right] \nonumber \\
  &&\qquad\qquad \times C(x,t;u,v;D,\mu) = 0
\label{calsym}
\end{eqnarray}
for the {\em renormalized} correlation function $C(x,t)$,
\begin{equation}
  \langle s(x,t) s(x',t') \rangle = C(x-x',t-t') \ .
\label{2pcorf}
\end{equation}
The standard solution of the partial differential equation (\ref{calsym})
proceeds via the method of characteristics, whereupon we parametrize 
$\mu(\ell) = \mu \ell$. For the correlation function, this leads to
\begin{eqnarray}
  &&C(x,t;u,v;D,\mu) = \exp 
  \left[ \int_1^\ell \gamma (\ell') \frac{d\ell'}{\ell'} \right] \nonumber \\ 
  &&\qquad\quad \times 
  C \left(x,t; u(\ell), v(\ell); D(\ell), \mu \ell \right)
\label{cssol1} \\
  &&\qquad = (\mu\ell)^{d-2} \exp 
  \left[ \int_1^\ell \gamma (\ell') \frac{d\ell'}{\ell'} \right] \nonumber \\ 
  &&\qquad\quad 
  \times {\hat C} 
  \left( \mu \ell x, (\mu\ell)^2 \, D(\ell) \, t; u(\ell), v(\ell) \right) \ .
\label{cssol2}
\end{eqnarray}
In the last step, we have explicitly extracted the scaling dimension of the
fields ($[s(\vec{x},t)] = \mu^{(d-2)/2}$), and thus rendered the quantity
${\hat C}$ dimensionless. The ``running couplings'' are furthermore given by
the solution of the flow equations
\begin{eqnarray}
  \ell \frac{d u(\ell)}{d\ell} = \beta_u \Bigl( u(\ell),v(\ell) \Bigr) \, , 
  \nonumber \\
  \ell \frac{d v(\ell)}{d\ell} = \beta_v \Bigl( u(\ell),v(\ell) \Bigr) \ ,
\label{floweq}
\end{eqnarray}
and
\begin{equation}
  \ell \frac{d D(\ell)}{d\ell} = 
  D(\ell) \, \zeta \Bigl( u(\ell),v(\ell) \Bigr) \ .
\label{flodeq}
\end{equation}

In the vicinity of a renormalization-group fixed point, Eq.~(\ref{cssol2}) 
reduces to
\begin{equation}
  C(x,t) \propto \ell^{d-2+\gamma^*} 
  {\hat C} \left(  \mu\ell \, x, \mu^2 D \, \ell^{2+\zeta^*} \, t;
                   u_*,v_* \right) \ ,
\label{csslfp}
\end{equation}
with $\gamma^* = \gamma(u_*,v_*)$ and $\zeta^* = \zeta(u_*,v_*)$. Upon 
employing the matching condition $\mu\ell \, x = 1$, one finally obtains the
identification of the scaling exponents with the above anomalous dimensions
$\gamma^*$ and $\zeta^*$ at a stable RG fixed point.

Using the conventional notation for dynamic critical phenomena, we write the
two-point correlation function in the scaling form
\begin{equation}
  C(x,t) = \frac{1}{x^{d-2+\eta}} \, {\hat C}(t / x^z) \ .
\label{corscx}
\end{equation}
Here, $\eta$ and $z$ are the {\em Wilson-Fisher} and {\em dynamic} scaling
exponents, respectively, and ${\hat C}(y)$ is a scaling function with the
asymptotic limits ${\hat C}(y) \to {\rm const.}$ for $y \to 0$, and 
${\hat C}(y) \to y^{(2-d-\eta)/z}$ for $y \to \infty$.
In Fourier space, Eq.~(\ref{corscx}) becomes
\begin{equation}
  C(q,\omega) = \frac{1}{q^{z+2-\eta}} \, {\hat C}(\omega / q^z) \ .
\label{corscq}
\end{equation}
In terms of the anomalous dimensions $\gamma^*$ and $\zeta^*$, according to
Eq.~(\ref{csslfp}) these two independent exponents read
\begin{equation}
  \eta = \gamma^* \ , \quad {\rm and} \quad z = 2 + \zeta^* \ .
\label{sclexp}
\end{equation}
In the context of non-equilibrium growth and interface scaling phenomena,
Eq.~(\ref{corscx}) is usually expressed in terms of the {\em roughness} 
exponent $\chi$ instead of $\eta$, 
\begin{eqnarray}
  C(x,t) = x^{2\chi} \, {\hat C}(t / x^z) \, .
\end{eqnarray}
Thus we identify
\begin{equation}
  \chi = \frac{2 - d - \eta}{2} = \frac{2 - d - \gamma^*}{2} \ .
\label{rghexp}
\end{equation}
For the Gaussian (Edwards-Wilkinson) model with $g = 0$, which corresponds to a
simple noisy diffusion equation, one immediately obtains $\eta = 0$, $z = 2$, 
and $\chi = 1 - \frac{d}{2}$.

We finally note that any {\em finite, non-zero} fixed point $0 < u_* < \infty$ 
implies $d-2 + \gamma^* - 2 \zeta^* = 0$, see (\ref{betag2}), and therefore the
non-trivial scaling relation
\begin{equation}
  z + \chi = 2 \ .
\label{screlg}
\end{equation}
In the same manner, Eq.~(\ref{betaf2}) leads to the additional {\em exact} 
result 
\begin{equation}
  z_{\rm lr} = \frac{4 + d - 2 \rho}{3} \ ,
\label{screlf}
\end{equation}
{\em provided} $0 < v_* < \infty$.  In situations where Eq.~(\ref{screlg}) 
holds as well, this already fixes the scaling exponents of the KPZ problem with
stable long-range correlated noise to
\begin{equation}
  \eta_{\rm lr} = \frac{2 - d - 4 \rho}{3} \ , \quad 
  \chi_{\rm lr} = \frac{2 - d + 2 \rho}{3} \ .
\label{scaexp}
\end{equation}
Furthermore, if $\beta_v$ is strictly positive, then $v \to 0$ under scale
transformations, and the short-range fixed point is {\em stable} against the
long-range noise correlations. In terms of the short-range dynamic exponent
$z_{\rm sr}$, Eq.~(\ref{betaf2}) thus implies the {\em exact} stability
condition
\begin{equation}
  z_{\rm sr} < z_{\rm lr} = \frac{4 + d - 2 \rho}{3}
\label{stacon}
\end{equation}
for the KPZ fixed point with local noise.

\begin{figure}[ht]
  \centerline{\epsfysize=0.6\columnwidth \epsfbox{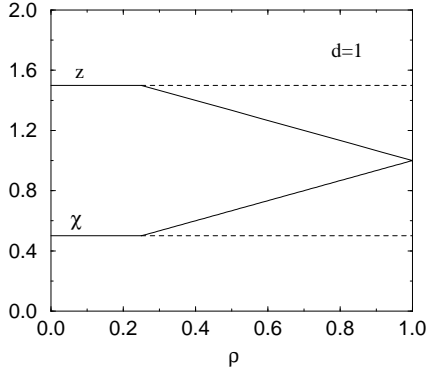}}
\caption{Exact roughness and dynamic exponent as a function of the exponent
  $\rho$ of the noise correlations in the ($1+1$)--dimensional model. Above a
  critical value $\rho_c=\frac{1}{4}$ the long-range fixed point becomes 
  stable.}
\label{exact_exponents_1.eps}
\end{figure}

In one dimension, and for purely short-range noise, i.e., $w = 0$, we know that
the stationary distribution is given by (\ref{eq:1dstat}). Thus we immediately
see that the statics is characterized by Gaussian exponents, and hence
\begin{equation}
  d = 1: \quad \eta_{\rm sr} = 0 \ , \quad \chi_{\rm sr} 
  = \frac{1}{2} \ , \quad 
  z_{\rm sr} = \frac{3}{2} \ ,
\label{1dsrex}
\end{equation}
where we have used the scaling relations (\ref{rghexp}) and (\ref{screlg}),
which is valid in the non-linear regime. Eq.~(\ref{stacon}) then yields the
stability condition for the short-range fixed point {\em to all orders} of
perturbation theory, namely
\begin{equation}
  \rho < \frac{1}{4}
  \ \Longrightarrow \ {\rm short-range \ fixed \ point \ stable} \ .
\label{1dstcd}
\end{equation}

The task of an explicit renormalization group analysis is therefore to 
establish the fixed-point structure of the field theory (\ref{dynfun}). It 
turns out that to this end one has to distinguish the cases $d < 2 (1+\rho)$
and $d \geq 2 (1+\rho)$. For, as can be already inferred from the naive scaling
dimension of the long-range coupling $v$, see Eqs.~(\ref{noiren}) and 
(\ref{betaf2}),
\begin{equation}
  d_c = 2 (1+\rho)
\label{lowcrd}
\end{equation}
defines a {\em lower} critical dimension for the roughening transition in the 
KPZ problem with long-range spatially correlated noise. In the following two
sections, we shall discuss the distinct scaling regimes below and above the 
lower critical dimension (\ref{lowcrd}) separately.

\section{Scaling behavior below the critical dimension} 
\label{1dren}

\subsection{Non--minimal renormalization at fixed dimension}
\label{non_minimal}

We start by evaluating the renormalization constants for {\em fixed} dimension
$d$ {\em and} correlation exponent $\rho$. The two independent $Z$ factors $Z$
and $Z_D$ can be determined by evaluating the two-point vertex functions
$\partial_{q^2} \Gamma_{{\tilde s} s}(\vec{q},\omega)$ and 
$\Gamma_{{\tilde s}{\tilde s}}(\vec{q},\omega)$ at the normalization point 
$\vec{q} = \vec{0}$, $\omega/ 2 D = \mu^2$. The corresponding one-loop diagrams
are identical to those for the conventional KPZ problem with short-range noise,
see e.g. Ref.~\cite{frey-taeuber:94}, and yield to leading order in the 
external momentum $\vec{q}$
\begin{eqnarray}
  &&\Gamma_{{\tilde s} s}(q,\omega) = i \omega + D q^2 \Biggl[
  1 - \frac{d-2}{4d} \, g^2 \int_p \frac{1}{p^2 + i \omega / 2D} \nonumber \\
  &&\qquad - \frac{d-2(1+\rho)}{4d} \, w g^2 \int_p 
  \frac{p^{-2\rho}}{p^2 + i\omega/2D} \Biggr] \ , 
\label{1lgats}
\end{eqnarray}
and
\begin{eqnarray}
  &&\Gamma_{{\tilde s}{\tilde s}}(q,\omega) = 
  -2 D \Biggl[ 1 + w q^{-2\rho} + \frac{g^2}{4} 
  \Re \int_p \frac{1}{p^2 + i\omega/2D} \nonumber \\
  &&\quad + \frac{w g^2}{2} 
  \Re \int_p \frac{p^{-2\rho}}{p^2 + i\omega/2D} + \frac{w^2 g^2}{4} 
  \Re \int_p \frac{p^{-4\rho}}{p^2 + i\omega/2D} \Biggr] \ . \nonumber \\ &&
\label{1lgatt} 
\end{eqnarray}
At the normalization point, and in dimensional regularization, the real parts 
of the above momentum integrals become
\begin{eqnarray}
  \Re &&\int_p \frac{1}{p^{2\sigma} (p^2 + i \mu^2)} 
  = - \, \frac{\Gamma(2+\sigma-\frac{d}{2}) \Gamma(\frac{d}{2}-\sigma)}
               {2^{d-1} \pi^{\frac{d}{2}} \Gamma(\frac{d}{2})} \nonumber \\
  &&\quad \quad \quad \times \frac{\mu^{d-2(1+\sigma)}}{d-2(1+\sigma)} 
          \ \sin \left[ \frac{\pi}{4} (d-2 \sigma) \right] \ .
\label{dimreg}
\end{eqnarray}
Eq.~(\ref{1lgats}) then leads to
\begin{equation}
  Z Z_D = 1 - \frac{1}{4} \ {A_d \, g_0^2 \, \mu^{d-2}}
  \left( 1 + w_0 \, \frac{B_d (\rho)}{A_d} \, \mu^{-2 \rho} \right) \ ,
\label{1lpzzd}
\end{equation}
with the $d$-dependent coefficients $A_d$ and $B_d(\rho)$ from 
Eqs. (\ref{geofcA}) and (\ref{geofcB}). The vertex function (\ref{1lgatt}) 
explicitly shows that the non-analytic contribution $\sim w_0 q^{-2\rho}$ is 
not renormalized by fluctuations. From Eq.~(\ref{1lgatt}) we find
\begin{eqnarray}
  Z_D = 1 &+& \frac{d \, A_d}{4 (d-2)} \, g_0^2 \, \mu^{d-2}
  \nonumber \\
  &+& \frac{d \, B_d (\rho)}{2 [d-2(1+\rho)]} \, w_0 g_0^2 \, \mu^{d-2(1+\rho)}
  \nonumber \\
  &+& \frac{d \, B_d (2\rho) }{4[d-2(1+2\rho)]}
      \, w_0^2 g_0^2  \, \mu^{d-2(1+2\rho)} \ ,
\label{1lpztd} 
\end{eqnarray}
and thus
\begin{eqnarray}
  Z = 1 &-& \frac{d-1}{2(d-2)} \, A_d \, g_0^2 \, \mu^{d-2}
  \nonumber \\
  &-& \frac{3d-2(1+\rho)}{4[d-2(1+\rho)]} \, B_d (\rho) 
      \, w_0 g_0^2 \, \mu^{d-2(1+\rho)}
  \nonumber \\
  &-&\frac{d\, B_d (2\rho)}{4[d-2(1+2\rho)]} 
     \, w_0^2 g_0^2  \, \mu^{d-2(1+2\rho)} \ .
\label{1lpzet}
\end{eqnarray}
Hence, there appear single poles in dimensional regularization at $d=2$,
$d=2(1+\rho)$, and $d=2(1+2\rho)$, above which the corresponding integrals
become ultraviolet-divergent. Notice, however, that the product $Z Z_D$ in
Eq.~(\ref{1lpzzd}) is non-singular. In a {\em minimal} renormalization scheme,
one would therefore set $Z Z_D = 1$ to one-loop order. While this is perfectly 
legitimate in the framework of an $\varepsilon$-expansion {\em above} the 
(lower) critical dimension, see Sec.~\ref{sec:rough}, this minimal subtraction
procedure would be inappropriate for $d < d_c$. This becomes already obvious in
the short-range noise case ($w_0 = 0$), where one would require that a 
fixed-dimension approach should recover the {\em exact} result 
$\eta_{\rm sr} = 0$ in one dimension. Yet, this is clearly impossible with a 
non-trivial renormalization constant $Z = Z_D^{-1} = 1 - d A_d g_0^2 \mu^{d-2} 
/ 4 (d-2)$. On the other hand, the term $\sim g_0^2$ vanishes at $d=1$ in the 
``full'' Z factor (\ref{1lpzet}), as desired. In order to correctly incorporate
the one-dimensional case into the theory, one has to apply a {\em non-minimal} 
renormalization scheme, and must retain not just the residues of the 
ultraviolet poles, but the full dimension dependence of the loop integrals 
\cite{frey-taeuber:94,taeuber-frey:95}. For the KPZ problem with local noise 
correlator, an explicit two-loop analysis on this basis was performed in 
Ref.~\cite{frey-taeuber:94}, which furthermore showed that the fixed-point 
value of the renormalized coupling $u$ diverges at the critical dimension 
$d_c = 2$. In light of additional arguments presented in Ref.~\cite{laessig:95}
that such a singularity is prone to emerge in any perturbational approach to 
the KPZ problem for $d \uparrow d_c$, this divergence clearly renders any 
attempt of an $\varepsilon$-expansion below $d_c$ obsolete.

Differentiation with respect to the normalization scale $\mu$, and expressing
the result in terms of the renormalized quantities, yields to first order in 
the couplings $u$ and $v$,
\begin{equation}
  \zeta = \frac{d-2}{4} \, u 
        + \frac{d-2(1+\rho)}{4} \, v \ ,
\label{1lgamd}
\end{equation}
and
\begin{equation}
  \gamma = - \frac{d-1}{2} \, u 
           - \frac{3d-2(1+\rho)}{4} \, v 
           - \frac{d}{4} \, h_d (\rho) \, \frac{v^2}{u} \ .
\label{1lgams}
\end{equation}
Here we have defined the function
\begin{eqnarray}
  h_d (\rho) &=& \frac{B_d (2 \rho) A_d}{B_d^2(\rho)} \nonumber \\
             &=& \frac{B(2-\frac{d}{2},\frac{d}{2}) \, 
                       B(2+2\rho-\frac{d}{2},\frac{d}{2}-2\rho)}
                      {[B(2+\rho-\frac{d}{2},\frac{d}{2}-\rho)]^2} 
             \nonumber \\
             &&\times  \frac{\sin\left[ \frac{\pi}{4} (d-4\rho)\right]
                       \sin\left( \frac{\pi}{4} d \right)}
                       {\sin^2 \left[ \frac{\pi}{4} (d-2\rho)\right]} \ ,
\end{eqnarray}
using Euler's beta function $B(x,y) = \Gamma(x) \Gamma(y)/ \Gamma(x+y)$.
Combining the above results finally yields the RG beta functions
\begin{eqnarray}
  \beta_u(u,v) 
  &=&(d-2) \, u - \frac{2d-3}{2} \, u^2 \nonumber \\
  &-& \frac{5d-6(1+\rho)}{4}  \, u v - \frac{d}{4} \, h_d (\rho)\, v^2 \ ,
\label{1lbetg} \\ 
  \beta_v(u,v) 
  &=& (d-2-2\rho) \, v - \frac{3}{4} (d-2) \, u v
  \nonumber \\ 
  &&- \frac{3}{4} (d-2-2\rho) \, v^2 \ .
\label{1lbetf}
\end{eqnarray}

\subsection{RG fixed points and scaling in one dimension}

\begin{figure}[htb]
  \centerline{\epsfxsize=0.7\columnwidth \epsfbox{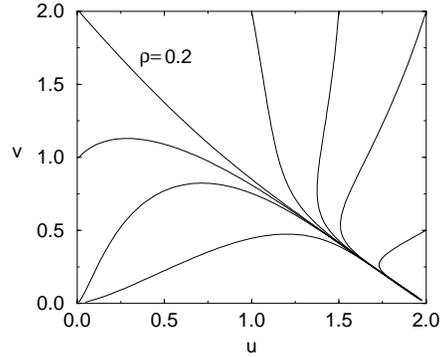}} 
\caption{Numerical solution of the RG flow equations (flow diagram) for the 
  one-dimensional KPZ equation with spatially long-range correlated noise
  characterized by an exponent $\rho = 0.2$: the short-range fixed point
  $(u_*,v_*) = (2,0)$ is stable.}
\label{fig:1dflow_02}
\end{figure}

The above one-loop beta functions (\ref{1lbetg}) and (\ref{1lbetf}) and ensuing
flow equations (\ref{floweq}) were essentially derived in 
Ref.~\cite{medina-hwa-kardar-zhang:89} already, with the slight modification
that there the function $h_d (\rho)$ was replaced with $h_d(0) = 1$.  This
minor change does not drastically alter the scaling behavior as function of
$\rho$ and $d$, however, and a discussion of the different regimes, albeit 
based on a one-loop approximation only, can be found in 
Ref.~\cite{medina-hwa-kardar-zhang:89}. Here, we thus restrict ourselves to the
analysis of the sole non-trivial {\em physical} realization below the critical
dimension, namely $d = 1$. We then have to distinguish between two different
situations:

(i) $0 \leq \rho < \frac{1}{4}$: In this case, apart from the Gaussian
(Edwards--Wilkinson) fixed point $u_* = 0$, $v_* = 0$, there exists only the
short-range KPZ fixed point
\begin{equation}
  u_* = 2 \ , \quad v_* = 0 \ ,
\label{1dsrfp}
\end{equation}
which furthermore is {\em stable} according to the exact result (\ref{1dstcd}).
The corresponding scaling exponents are those of (\ref{1dsrex}). A numerical 
solution of the flow equations (\ref{floweq}) for $\rho = 0.2$ is shown in 
Fig.~\ref{fig:1dflow_02}.

\begin{figure}[htb]
  \centerline{\epsfxsize=0.7\columnwidth \epsfbox{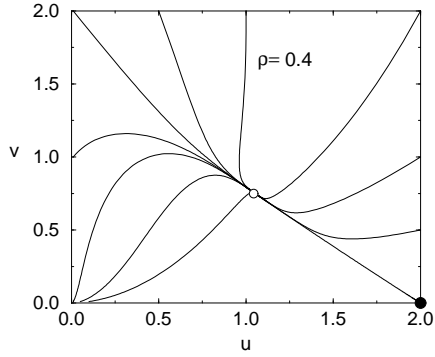}} 
\caption{Numerical solution of the RG flow equations (flow diagram) for the 
  one-dimensional KPZ equation with spatially long-range correlated noise
  characterized by an exponent $\rho = 0.4$: The short-range fixed point
  becomes unstable, and the stable long-range fixed point $(u_*,v_*)$ governs
  the asymptotic scaling behavior.}
\label{fig:1dflow_04}
\end{figure}

(ii) $\frac{1}{4} \leq \rho < 1$: In this regime, the short-range fixed point
(\ref{1dsrfp}) is unstable. There exists, however, also a non-trivial zero of
the beta functions (\ref{1lbetg}) and (\ref{1lbetf}), which leads to a 
quadratic equation with the solution
\begin{eqnarray}
  &&u_* = \frac{4}{3} \, \frac{1 + 2\rho} {h_1(\rho) - (1 - 4\rho^2)} 
  \Biggl[h_1(\rho) - (1-\rho-6\rho^2) \nonumber \\
  &&\quad + \sqrt{(1-\rho-6\rho^2)^2 - (1-2\rho-8\rho^2) \, h_1(\rho)} \Biggr]
  \; ,
\label{1dlrfg}
\end{eqnarray}
and
\begin{equation}
  v_* = \frac{4}{3} - \frac{u_*}{1 + 2\rho} \ .
\label{1dlrff}
\end{equation}
The non-trivial long-range fixed point is stable, as can be seen in the flow
diagram for the case $\rho = 0.4$, which is depicted in 
Fig.~\ref{fig:1dflow_04}. From (\ref{screlf}) and (\ref{scaexp}) we therefore 
infer the values for the long-range noise scaling exponents in one dimension,
\begin{equation}
  \eta_{\rm lr} = \frac{1-4\rho}{3} < 0 \ , \quad 
  \chi_{\rm lr} = \frac{1+2\rho}{3} \ , 
\label{lrexps}
\end{equation}
and
\begin{equation}
  z_{\rm lr} = \frac{5 - 2\rho}{3} \ .
\label{lrexpz}
\end{equation}

(iii) $\rho > 1$: For $\rho > 1$ the above exact results for the dynamic 
exponent become invalid, since $z$ becomes smaller than $1$, which would imply
motion faster than ballistic transport. Surprisingly, however, Hayot and
Jayaprakash \cite{hayot-jayaprakash:96,hayot-jayaprakash:97} find agreement
with the exact result for the roughness exponent $\chi$ even for $\rho \geq 1$.
This interesting observation still lacks a deeper theoretical understanding.

Thus, despite the uncontrolled non-minimal renormalization prescription which
does not allow for the identification of a small expansion parameter in the
perturbation series, we could obtain the {\em exact} results (\ref{1dsrex}) and
(\ref{lrexps}) for the scaling exponents at the short-range and long-range
fixed points, respectively, as well as for the stability criterion 
(\ref{1dstcd}). In the short-range case, the ``hidden'' fluctuation-dissipation
theorem in one dimension, and the scaling relation (\ref{screlg}) are 
sufficient to fix the scaling exponents. For power-law correlated noise, which
cannot be renormalized perturbatively, the additional condition of a non-zero, 
finite fixed point $v_*$ determined $z$, and via Eq.~(\ref{screlg}) $\chi$ as 
well. This remarkable fact is also intimately related to the mapping of the 
noisy Burgers equation to a driven diffusive system in one dimension. As 
opposed to the KPZ problem, driven diffusive systems do permit a systematic 
$\varepsilon$ expansion {\em below} $d_c = 2$, and are therefore subject to a 
controlled perturbation expansion \cite{janssen-schmittmann:86}. A direct 
analysis of driven diffusive systems with long-range correlated spatial noise 
in fact yields the identical stability condition (\ref{1dstcd}) for the 
short-range fixed point, and confirms the result (\ref{lrexps}) 
\cite{janssen-schmittmann:98}.

\section{The roughening transition and scaling above the critical dimension} 
\label{sec:rough}

Above the critical dimension $d_c = 2(1+\rho)$, a systematic perturbational
expansion in terms of $d-d_c$ is possible within a {\em minimal}
renormalization scheme. As was demonstrated to two-loop order in
Ref.~\cite{frey-taeuber:94}, the short-range noise KPZ equation has a
non-trivial {\em unstable} fixed point for $d \geq d_c= 2$, which vanishes as
$d \to 2$ (see also Ref.~\cite{wiese:97}). It is therefore possible to design a
controlled $(2 + \varepsilon)$-expansion with a well-defined small parameter, 
as opposed to the situation below the critical dimension. The unstable fixed 
point describes a {\em non-equilibrium roughening transition}, separating a 
smooth phase with Edwards--Wilkinson scaling exponents from a strong-coupling 
rough phase, which remains inaccessible through perturbational means
\cite{laessig:95}. The $(2 + \varepsilon)$-expansion itself works in very close
analogy with the standard field-theoretic treatment of the non-linear sigma 
model, where the unstable fixed point corresponds to the critical temperature 
(see Appendix \ref{appb}). Upon employing the directed-polymer representation 
of the KPZ problem, it was subsequently demonstrated that the one-loop beta 
function obtained within such an $\varepsilon$-expansion is actually 
{\em exact} to all orders in perturbation theory 
\cite{laessig:95,janssen_unpub:97,wiese:98}.

The KPZ problem with power-law correlated noise can be treated in much the same
manner. As in other problems where non-analytic long-range expressions are
present in the bare vertices, the expansion about the critical dimension has to
be supplemented with an expansion with respect to the exponent $\rho$
\cite{sak:73,gusmao-theumann:83,theumann-gusmao:85,honkonen-nalimov:89}.
Correspondingly, we may construct a minimal renormalization scheme, where the
residues of the $\varepsilon = d-2$ poles in the $Z$ factors are to be taken at
$d = 2$ and $\rho = 0$. For our previous one-loop results, this immediately
means that
\begin{equation}
  Z \, Z_D = 1 \ ,
\label{minzzd}
\end{equation}
and therefore
\begin{equation}
  \zeta = 0 \ , \quad {\rm and} \ \gamma = -\, \frac{(u + v)^2}{2 u}
\label{minren}
\end{equation}
in minimal subtraction. Notice that the above result for $\zeta$ implies $z=2$ 
at {\em any} RG fixed point, see Eq.~(\ref{sclexp}). Furthermore, 
Eqs.~(\ref{betag2}) and (\ref{betaf2}) yield the one-loop RG beta functions
\begin{eqnarray}
  \beta_u(u,v) &=& \varepsilon \, u - \frac{1}{2} \, (u + v)^2 \ ,
\label{mnbetg} \\
  \beta_v(u,v) &=& (\varepsilon - 2 \rho) \, v \ .
\label{mnbetf}
\end{eqnarray}
The same expressions may of course be obtained by setting $d = 2$ and $\rho =
0$ in the quartic terms in (\ref{1lbetg}) and (\ref{1lbetf}).

Notice that by using an expansion with respect to $\rho$, we restrict ourselves
to fixed points which are of the same type as the fixed points at $\rho = 0$, 
i.e., fixed points describing the smooth phase and the roughening transition
of the short-range problem. We cannot expect to learn anything about the 
strong-coupling regime this way. In general the fixed-point structure may 
radically change as a function of $\rho$. In fact, this must be the case for 
the KPZ equation, as may be inferred from considering the particular case 
$\rho = 1$, which corresponds to the Burgers equation with non-conserved noise 
(``model B'' in the notation of Forster {\em et\ al.} 
\cite{forster-nelson-stephen:77}). The latter model displays no non-equilibrium
phase transition, and is known to have an {\em upper} critical dimension 
$d_c=4$. A straightforward renormalization-group analysis of the Burgers 
equation with non-conserved noise \cite{forster-nelson-stephen:77,zhang:86} 
tells us that the dynamic exponent is {\em exactly} given by $z=(2+d)/3$, and 
that there is a stable fixed point $v_*$ to order $4-d$. From our general 
analysis in section \ref{gener} (which is valid to any order in perturbation 
theory), we also know that this exponent is the analytic continuation of the 
strong-coupling (long-range noise) fixed point. By using an expansion with 
respect to $\rho$, this fixed point will obviously {\em not} be accessible. 
Hence, there must be some borderline in the $(d,\rho)$--plane where a
non-analytic change in the fixed-point structure takes place. We shall return 
to this issue in our final section \ref{discs}. For now, we restrict ourselves 
mainly to a discussion of the roughening transition above the critical 
dimension $d_c = 2 (1 + \rho)$.

\subsection{Cole-Hopf transformation and minimal renormalization above the
  critical dimension}
\label{sec:cole_hopf}

We will show in this section that, similar to the short-range case 
\cite{laessig:95}, the above one-loop RG beta functions actually hold to 
{\em all orders} in perturbation theory. The simplest way to see this is via a
stochastic Cole--Hopf transformation \cite{janssen_unpub:97} (see also
Refs.~\cite{grassberger:82,wiese:98}), which maps the KPZ equation onto the
problem of directed polymers in a random environment
\cite{huse-henley-fisher:85,kardar-zhang:87,fisher-huse:91,hwa-fisher:94}.
Hence we define new fields $n(\vec{x},t)$ via
\begin{equation}
  n(\vec{x},t) = \frac{2}{g} \, \exp \left[ \frac{g}{2} \, 
  s(\vec{x},t) \right] \ .
\label{holkfs}
\end{equation}
In the discretized form of the dynamic functional resulting from 
(\ref{eq:prob_distribution}), this highly non-linear transformation is to be 
supplemented with
\begin{equation}
  {\tilde n}(\vec{x},t+\tau) = {\tilde s}(\vec{x},t+\tau) \, 
  \exp \left[ - \frac{g}{2} \, s(\vec{x},t) \right] \ ,
\label{holkts}
\end{equation}
which renders the Cole--Hopf transformation from the fields $(s,{\tilde s})$ to
$(n,{\tilde n})$ {\em pseudo-canonical}, i.e., with Jacobian 
\begin{equation}
  \frac{\partial (n,{\tilde n})}{\partial (s,{\tilde s})} = 1 \, , \quad 
  {\rm and} \quad {\tilde s} \, \frac{\partial s}{\partial t} 
  = {\tilde n} \, \frac{\partial n}{\partial t} \ ,
\label{trterm}
\end{equation}
as long as one neglects the time increment $\tau$, and considers the fields as
analytic functions. However, we need to be slightly more careful here since it
is known \cite{janssen:92} that the time derivatives of the fields as well as
the response fields are actually {\em nowhere} continuous. Thus, by means of 
such a non-linear transformation, one has to take recourse to the discretized 
form of the functional, or alternatively use different methods like 
Fokker-Planck equations or Langevin equations in a well defined form 
\cite{risken:88}. Then one finds that the apparently trivial terms in 
Eqs.~(\ref{trterm}) produce a non-straightforward contribution 
$-D g R\left( 0\right) \tilde{n}n / 2$ to the integrand of the dynamic 
functional (\ref{eq:dyn_functional}), which is proportional to the noise.
Furthermore,
\begin{equation}
  \nabla \tilde{n}\nabla n=\nabla \tilde{s}\nabla s
  -g\tilde{s}\left( \nabla s\right) ^{2}/2 \ ,
\end{equation}
which eliminates the non-linear term in the effective equation of motion,
leaving only the noise vertex. The dynamic functional in terms of the fields
$n$ and $\tilde{n}$ thus becomes
\begin{eqnarray}
  &&{\cal J} = \int dt \Biggl\{ \int_x \tilde{n}({\bf x},t) 
  \Bigl[ \dot{n}({\bf x},t) -D\nabla^{2} n({\bf x},t) \nonumber \\
  &&\qquad \qquad \qquad -\frac{Dg}{2} R(0) n({\bf x},t) \Bigr]
\label{19} \\
  &&\quad -\frac{Dg^{2}}{4}\int_{x} \int_{x'} \tilde{n}({\bf x},t)
  n({\bf x},t) R( {\bf x}-{\bf x}^{\prime}) \tilde{n}({\bf x}^{\prime },t)
  n({\bf x}^{\prime },t) \Biggr\} \ .  \nonumber 
\end{eqnarray}
Now it becomes obvious that the seemingly problematic term $\propto R(0)$ can 
be eliminated by introducing an additional term in the stochastic Cole--Hopf 
transformation: one merely has to replace $s({\bf x},t) \rightarrow 
s({\bf x},t) +D R(0)t$ in the exponentials of Eqs.~(\ref{holkfs}) and
(\ref{holkts}). This justifies our dropping this term in the following, 
whereupon we write the dynamic functional in terms of the fields $n$ and
${\tilde n}$ in the form
\begin{eqnarray}
  {\cal J} &=& \int \! d^dx \int \! dt \ \Biggl[ {\tilde n} 
  \left( \partial_t n - D \vec{\nabla}^2 n \right) \nonumber \\
  &&- \frac{D g^2}{4} \left( [{\tilde n} n]^2 + 
    w \left[ \vec{\nabla}^{-\rho} ({\tilde n} n) \right]^2 \right) \Biggr] \ .
\label{holkac}
\end{eqnarray}

\begin{figure}
  \centerline{\epsfxsize 0.6\columnwidth \epsfbox{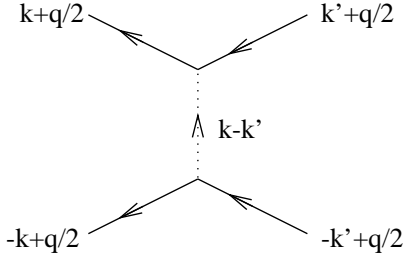}} 
\caption{Four-point vertex for the KPZ problem with spatially long-range 
  correlated noise in the Cole--Hopf representation.}
\label{vertex}
\end{figure}

\begin{figure}
  \centerline{\epsfxsize 0.9\columnwidth \epsfbox{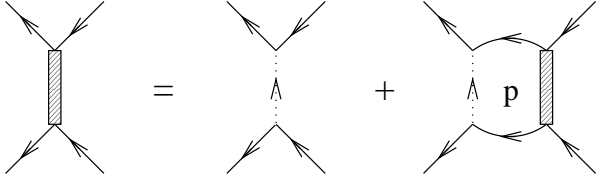}} 
\caption{Diagrammatic representation of the Bethe--Salpeter equation for the
  four-point noise vertex.}
\label{betsal}
\end{figure}

When we introduce renormalized quantities according to
Eqs.~(\ref{fldren})--(\ref{noiren}), and furthermore demand that the relations
(\ref{holkfs}) and (\ref{holkts}) hold for the renormalized fields as well, we
find
\begin{equation}
  {\tilde n} \, n \rightarrow Z \, Z_D \, {\tilde n} \, n \ ,
\label{holkfr}
\end{equation}
and the renormalized action reads
\begin{eqnarray}
  &&{\cal J} = \int \! d^dx \int \! dt \Biggl[ Z Z_D \, {\tilde n}
  \left( \partial_t n - Z Z_D \, D \vec{\nabla}^2 n \right) 
\label{holkra} \\
  &&- \frac{D \mu^{2-d}}{4} \left( Z_D \, u A_d^{-1} ({\tilde n} n)^2 + 
    v B_d^{-1} \mu^{2\rho} \left[ \vec{\nabla}^{-\rho} ({\tilde n} n) \right]^2
    \right) \Biggr] \ . \nonumber
\end{eqnarray}
However, in the field theory (\ref{holkac}) with the bare diffusion propagator
$(-i \omega + D q^2)^{-1}$ and the four-point noise vertex depicted in
Fig.~\ref{vertex}, there exist {\em no loop diagrams} contributing to the
renormalization of the propagator. Therefore, in a {\em minimal} 
renormalization scheme, Eq.~(\ref{minzzd}), and hence Eq.~(\ref{mnbetf}) hold 
to {\em all} orders in perturbation theory.
This leaves us with the determination of $Z_D$ from the renormalization of the
noise vertex.  As a consequence of the absence of any propagator
renormalizations, this is readily achieved via a summation of the entire
perturbation series by means of a {\em Bethe--Salpeter equation}, see
Fig.~\ref{betsal}.  
This is actually in complete analogy with the field-theoretic analysis of 
diffusion-limited pair annihilation reactions, see 
Ref.~\cite{peliti:84,lee:94,lee-cardy:95}. Upon defining
\begin{eqnarray}
  &&\Gamma_{22} \left( \vec{k},\vec{k}^\prime; 
                       \frac{i\omega}{2D} + \frac{q^2}{4} \right) =
  - \Gamma_{{\tilde n}{\tilde n}nn} \bigg(-\frac{\vec{q}}{2}-\vec{k},
  -\frac{\omega}{2}; \nonumber \\
  &&\qquad -\frac{\vec{q}}{2}+\vec{k},-\frac{\omega}{2}; 
  \frac{\vec{q}}{2}+\vec{k}^\prime, \frac{\omega}{2}; 
  \frac{\vec{q}}{2}-\vec{k}^\prime, \frac{\omega}{2} \biggr) \, ,
\label{verfun}
\end{eqnarray}
the corresponding analytic expression, taken at the normalization point
$i\omega / 2D + q^2 / 4  = \mu^2$, reads
\begin{eqnarray}
  &&\Gamma_{22}(\vec{k},\vec{k}^\prime;\mu^2) = D g^2 \,
  R(\vec{k} - \vec{k}^\prime )  
\nonumber \\
  &&\qquad \quad + \frac{g^2}{4} \int_p \frac{R (\vec{k} - \vec{p})}
  {\mu^2 + p^2} \, \Gamma_{22}(\vec{p},\vec{k}^\prime;\mu^2) \ ,
\label{bsaleq} 
\end{eqnarray}
where $ R( \vec{k} ) = 1 + w k^{-2 \rho}$. 

In order to determine the renormalization factors from the Bethe--Salpeter
equation, it suffices to consider it at $\vec{k}^\prime = \vec{0}$ and to
leading order in the external momentum $\vec{k}$. Then the Bethe--Salpeter 
equation reduces to
\begin{eqnarray}
  &&\Gamma_{22}(\vec{k},\vec{0};\mu^2) = 
  D g^2 \, (1 + w k^{-2\rho}) \nonumber \\ 
  &&\qquad + \frac{g^2}{4} \int_p \frac {1 + w p^{-2\rho}}{\mu^2 + p^2} \, 
  \Gamma_{22}(\vec{p},\vec{0};\mu^2) \ .
\end{eqnarray}
Upon introducing $Z$ factors through Eqs.~(\ref{difren}), (\ref{verren}), and 
(\ref{noiren}), taking into account the identity (\ref{minzzd}) and requiring 
that to leading order in the momentum the renormalized vertex function assumes 
the form 
\begin{equation}
  \Gamma_{22}(\vec{k},\vec{0};\mu^2) = 
  D \mu^{2-d} \, \left[ u A_d^{-1} + v B_d^{-1} (\mu/k)^{2\rho} \right] \ ,
\end{equation}
one finds
\begin{eqnarray}
  &&u A_d^{-1} \mu^{2-d} \left[ 1 - \frac{g_0^2}{4} 
    \int_p \frac{1 + w_0 p^{-2\rho}}{p^2 + \mu^2} \right] \nonumber \\
  &&\qquad = g_0^2 \left[ 1 + \frac{w_0 g_0^2}{4}
    \int_p \frac{1 + w_0 p^{-2\rho}}{p^{2\rho} (p^2 + \mu^2)} \right] \ .
\label{bsalrn}
\end{eqnarray}
Evaluating the integrals with the aid of the formula (\ref{dimreg}), we thus
arrive at the {\em exact} result
\begin{eqnarray}
  &&Z_D = \Biggl( 1 + \frac{d A_d g_0^2 \mu^{d-2}}{4 (d-2)} + 
  \frac{d B_d w_0 g_0^2 \mu^{d-2(1+\rho)}}{4 [d-2(1+\rho)]} \Biggr) \nonumber\\
  &&\qquad \times \Biggl( 
  1 - \frac{d B_d w_0 g_0^2 \mu^{d-2(1+\rho)}}{4 [d-2(1+\rho)]} 
\label{mnzdur} \\
  &&- \frac{\Gamma(2+2\rho-\frac{d}{2}) \Gamma(\frac{d}{2}-2\rho)} 
  {2^{d-1}\pi^{\frac{d}{2}}\Gamma(\frac{d}{2})}
  \frac{w_0^2 g_0^2 \mu^{d-2(1+2\rho)}}{4 [d-2(1+2\rho)]} \Biggr)^{-1} \ .
  \nonumber
\end{eqnarray}
When this is expanded with respect to first order in the fluctuation
contributions, the one-loop result (\ref{1lpztd}) is recovered. In the minimal 
renormalization scheme, however, we have to set $d = 2 + \varepsilon$ and 
$\rho = 0$. The somewhat cumbersome expression (\ref{mnzdur}) then becomes
nicer when $Z_D$ is re-expressed in terms of the renormalized couplings,
\begin{equation}
  Z_D = \frac{1 + \frac{1}{2 \varepsilon} \, v 
          + \frac{1}{2 \varepsilon} \, \frac{v^2}{u}}
       {1 - \frac{1}{2 \varepsilon} \, u - \frac{1}{2 \varepsilon} \, v} \ .
\label{mnzdrn}
\end{equation}
In a similar manner, differentiating Eq.~(\ref{mnzdur}) with respect to the
normalization scale $\mu$, and expressing the result in terms of $u$ and $v$
yields precisely the RG beta functions (\ref{mnbetg}) and (\ref{mnbetf}), which
are thus valid to all orders in minimally renormalized perturbation theory.

Alternatively, one may utilize the structure of the perturbation expansion,
combined with the minimal renormalization prescription, to demonstrate the
validity of the beta functions (\ref{mnbetg}) and (\ref{mnbetf}) to all orders
in the perturbation series. In the following, we employ a {\em double} 
expansion in $\varepsilon$ and $\rho$, where it is assumed that both quantities
are of equal order of magnitude, in combination with minimal renormalization. 
Consider the perturbational expansion of the Bethe--Salpeter equation 
(\ref{betsal}) with respect to $g^{2}\propto u$ and $w$. A contribution 
proportional to $g^{2n}w^{k}$ ($k\leq n$) is accompanied by a momentum integral
that diverges superficially as $\Lambda ^{n\varepsilon -k\rho }$, where 
$\Lambda $ denotes an ultraviolet momentum cutoff. First, let $\varepsilon <0$.
Then, as shown by Honkonen and Nalimov \cite{honkonen-nalimov:89} in the 
analogous case of the Landau--Ginzburg--Wilson hamiltonian with long-range 
interactions, the contributions are effectively regularized by a combination of
the dimensional and the analytic regularization schemes. Instead of the 
UV divergen\-ces, now pole singularities of the combinations 
$\delta =-n\varepsilon +2k\rho $ arise with all possible positive integers $n$ 
and $k$. These poles are absorbed by the counterterms $(Z_{D}-1) u$. Thus, the 
general form of the renormalization factor $Z_D$ reads
\begin{eqnarray}
  Z_{D} &=&1+Y^{\left( 1\right) }+O\left( \delta ^{-2}\right) \ ,  \nonumber \\
  Y^{\left( 1\right) } &=&\sum_{n=1}^{\infty }\sum_{k=0}^{n+1} {n+1 \choose k}
  \frac{nC_{n,k}}{-n\varepsilon +2k\rho } \, u^{n} w^{k} \ .  
\label{21}
\end{eqnarray}
In general, the coefficients $C_{n,k}$ may be functions of $\varepsilon $ and
$\rho $, depending on the renormalization description. In the minimal
renormalization scheme, these coefficients are defined to be independent of
$\varepsilon $ and $\rho $. This procedure cancels all simple pole 
singularities arising in the limit $\varepsilon ,\rho \rightarrow 0$ as long as
$\rho $ is at least of the same order as $\varepsilon $. One can readily 
compute these coefficients, which are now independent of $\varepsilon $ and 
$\rho $, from the model without the correlated noise in the following way. 
First, consider the Bethe--Salpeter equation (\ref{betsal}) for $w=0$, i.e.\ 
with short-range correlation only. One can easily show that in this case 
$\left. \Gamma _{2,2}\right| _{w=0}=:\bar{\Gamma}_{2,2}$ is independent of 
${\bf k}$ and ${\bf k}^{\prime },$ and the solution of the Bethe--Salpeter 
equation (\ref{betsal}) is given by
\begin{equation}
  \bar{\Gamma}_{2,2}=\frac{-Dg^{2}}{\bar{Z}_{D}^{-1}-\frac{g^{2}}{4}
  \int_p \left( p^{2}+\mu ^{2}\right) ^{-1}} \ ,
\label{22}
\end{equation}
which is minimally renormalized exactly by means of
\begin{equation}
  \bar{Z}_{D}=\frac{1}{1-u/2\varepsilon } \ ,
\label{23}
\end{equation}
where $g^{2}=A_{d}^{-1} u\mu^{-\varepsilon }$. Now take the Bethe-Salpeter
equation (\ref{betsal}) in the limit $\rho \rightarrow 0$. This case can be
mapped to the preceding one with $w=0$ by setting $u\rightarrow \bar{u}=(1+w)u$
and
\begin{equation}
  \left( \left. Z_{D}\left( u,w\right) \right| _{\rho =0}+w\right) 
  u=\bar{Z}_{D}\left( \bar{u}\right) \bar{u} \ .  
\label{24}
\end{equation}
From this equation we get 
\begin{equation}
  \left. Z_{D}(u,w) \right|_{\rho =0}=1+\frac{(1+w)^{2}u}{2\varepsilon }
  +O\left( \varepsilon ^{-2}\right) \ ,  \label{25}
\end{equation}
from which we infer via comparison with the expansion (\ref{21}), and from
$C_{n,k}={\rm const.}$ for $\rho \neq 0$,
\begin{equation}
  Z_{D}\left( u,w\right) =1+\frac{u}{2}\left( \frac{1}{\varepsilon }
  +\frac{2w}{\varepsilon -2\rho }+\frac{w^{2}}{\varepsilon -4\rho }\right) 
  +O\left(\delta ^{-2}\right) \ .  
\label{26}
\end{equation}
The calculation of the logarithmic derivatives of the $Z$ factors requires only
the simple-pole contributions displayed in Eq.~(\ref{26}). With $Z Z_{D}=1$ we 
eventually arrive at
\begin{eqnarray}
  &&\gamma  = -\gamma _{D}=-\frac{1}{2}\left( 1+w\right) ^{2}u \; , \quad 
  \zeta = -\gamma -\gamma _{D}=0 \ ,  \nonumber \\
  &&\beta _{u} = \left( \varepsilon +\gamma \right) u \; , \quad 
  \beta _{w} = -\left( 2\rho +\gamma \right) w \ .  \label{27}
\end{eqnarray}
Note that these renormalization-group functions are {\em exact} within the 
minimal renormalization scheme.

\subsection{Scaling behavior in the smooth phase and at the roughening 
  transition}

We are now in a position to discuss the flow equations (\ref{floweq}) with the
minimally renormalized RG beta functions (\ref{mnbetg}) and (\ref{mnbetf}), 
valid to all orders in perturbation theory with respect to the non-linear 
couplings $u$ and $v$.  

\subsubsection{Exact results and flow diagrams}

\begin{figure}[htb]
\centerline{\epsfxsize 0.75\columnwidth \epsfbox{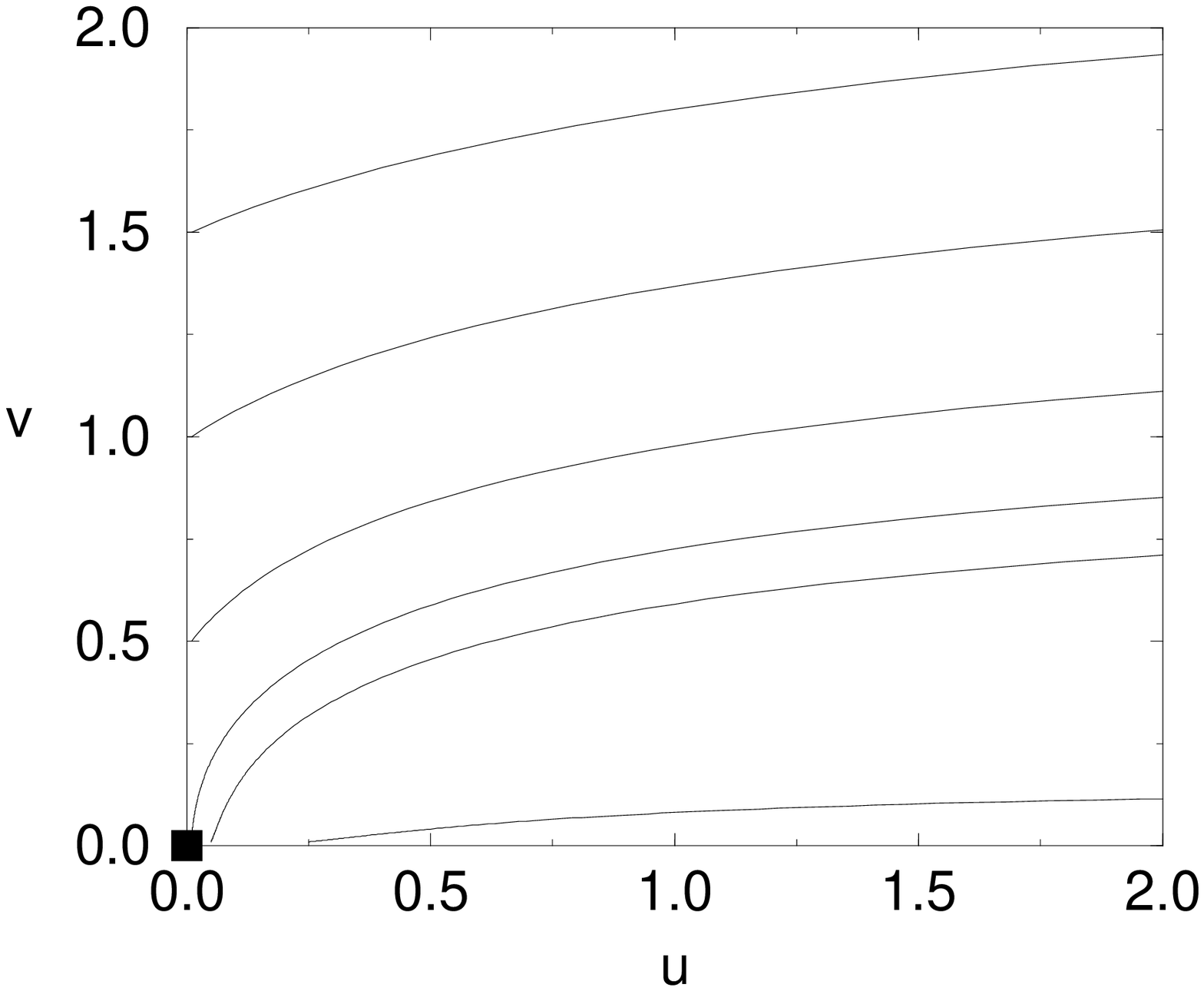}}
\centerline{\epsfxsize 0.75\columnwidth \epsfbox{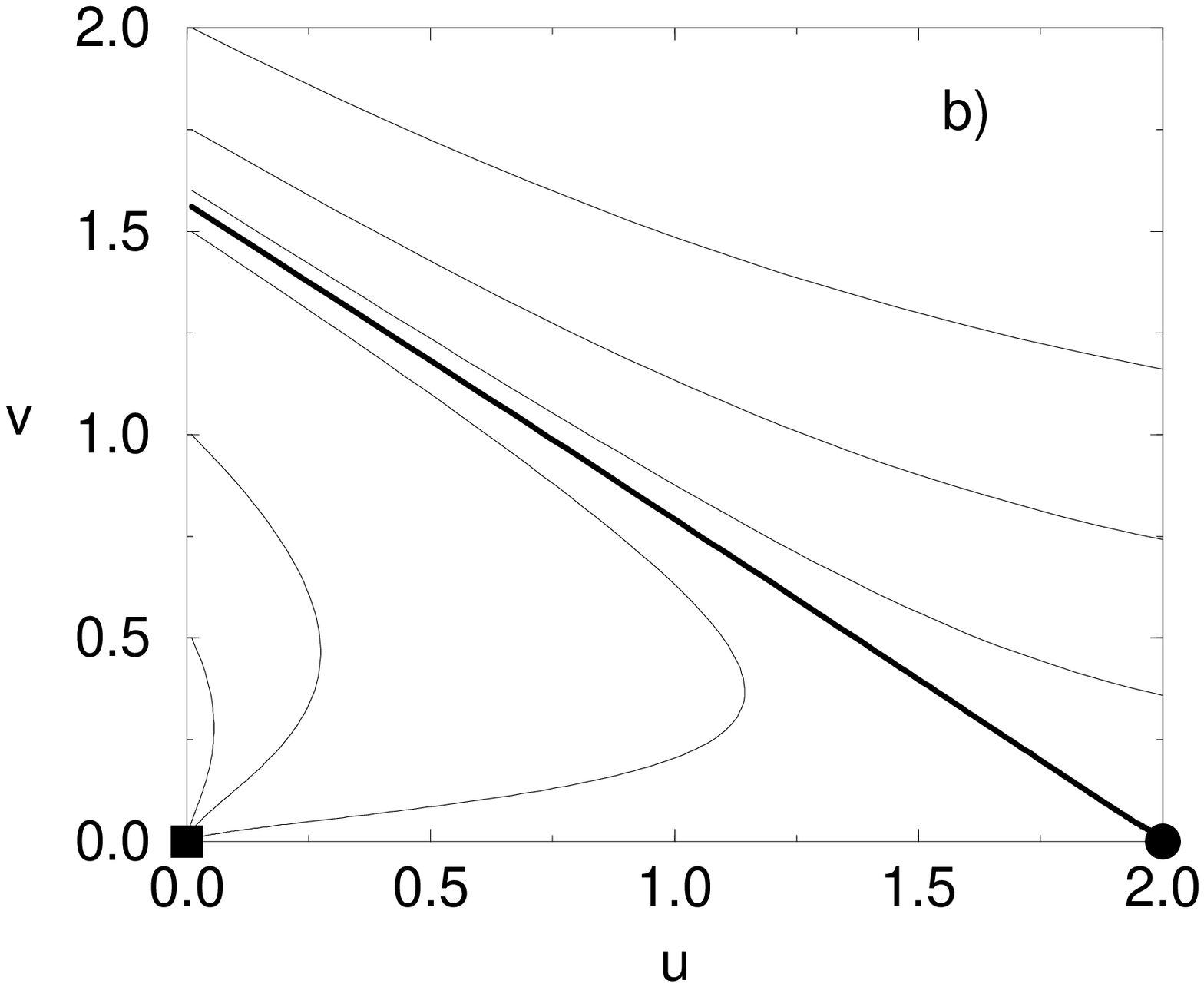}}
\caption{Flow diagram for the KPZ equation with spatially long-range correlated
  noise for (a) $d = 2$ and $\rho = 0.2$: the RG flow for $v$ diverges, and 
  perturbation theory breaks down; (b) $d = 3$ and $\rho = 0.2$: the long-range
  coupling $v \to 0$ under scale transformations, and the unstable short-range 
  fixed point $u_c = 2 \varepsilon$ separates a smooth phase ($u \to 0$) from a
  strong-coupling rough phase ($u \to \infty$).}
\label{mnflow}
\end{figure}

For $d < d_c = 2(1+\rho)$, i.e., $\varepsilon = d-2 < 2\rho$, 
Eq.~(\ref{mnbetf}) implies that $v(\ell) \to \infty$ as $\ell \to 0$, see 
Fig.~\ref{mnflow}(a). Thus in this situation, {\em minimally renormalized} 
perturbation theory (in the representation using the fields $n$ and 
${\tilde n}$) breaks down, and one must resort to the fixed-dimension expansion
discussed above, or an expansion near some other special point in the 
$(\rho,d)$--plane. For instance, in the particular case $\rho = 1$, which
represents the Burgers equation with non-conserved noise, an $\epsilon = 4-d$ 
expansion in the vicinity of the upper critical dimension $4$ can be
constructed \cite{forster-nelson-stephen:77}. 

On the other hand, for $d > d_c = 2(1+\rho)$, the RG flow equation shows that
$v(\ell) \to 0$ in the asymptotic infrared limit $\ell \to 0$. Hence the usual
short-range noise KPZ scenario comes into play
\cite{forster-nelson-stephen:77,kardar-parisi-zhang:86,%
  medina-hwa-kardar-zhang:89,frey-taeuber:94}. As depicted in 
Fig.~\ref{mnflow}(b), in addition to the trivial fixed point $u = 0$ and the
perturbatively non-accessible strong-coupling fixed point $u_* = \infty$,
there appears an {\em unstable} non-trivial fixed point
\begin{equation}
  u_c = 2 \, \varepsilon \ , \quad v_c = 0 \ .
\label{critfp}
\end{equation}
There is now a separatrix $v_c(u)$, depicted as a thick solid line in
Fig.~\ref{mnflow}(b), which marks the location of a {\em non-equilibrium}
phase transition. Eq.~(\ref{minren}) quite generally implies $z = 2$ in the 
minimal renormalization scheme, and furthermore the existence of the 
$O(\varepsilon)$ fixed point (\ref{critfp}) guarantees the validity of the 
scaling relation (\ref{screlg}). Thus, we find the following {\em critical} 
exponents at the roughening transition, valid to {\em all} orders in 
perturbation theory
\cite{doty-kosterlitz:92,laessig:95,janssen:97}
\begin{equation}
  \eta_c = 2 - d < 0 \ , \quad \chi_c = 0 \ , \quad z_c = 2 \ . 
\label{critex}
\end{equation}

Furthermore, we evaluate the stability matrix
\begin{eqnarray}
  \Omega(u,v) &=& \left( \begin{array}{cc} 
      \partial \beta_u / \partial u &\quad \partial \beta_u / \partial v \\
      \partial \beta_v / \partial u &\quad \partial \beta_v / \partial v \\
    \end{array} \right) \nonumber \\
  &=& \left( \begin{array}{cc} 
    \varepsilon - (u + v) &\quad - (u + v) \\ 0 &\quad \varepsilon - 2 \rho \\
  \end{array} \right) 
\label{stamat}
\end{eqnarray}
at the critical fixed point (\ref{critfp}).  For $\Omega(2\varepsilon,0)$, we
find the negative eigenvalue $- \phi_c$, which defines a {\em crossover
  exponent}
\begin{equation}
  \phi_c = \varepsilon = d - 2
\label{crcrex}
\end{equation}
or inverse {\em correlation length exponent} $\nu_c = \phi_c^{-1}$ associated
with the second-order dynamic phase transition at $u_c = g_c^2$
\cite{frey-taeuber:94}. The second eigenvalue of $\Omega(2\varepsilon,0)$ is
positive,
\begin{equation}
  \omega_c = \varepsilon - 2 \rho = d - 2(1+\rho) \ ,
\label{crscex}
\end{equation}
and can be interpreted as a {\em correction-to-scaling} exponent associated
with the {\em irrelevant} coupling $v$. Again, Eqs.~(\ref{crcrex}) and
(\ref{crscex}) constitute {\em exact} results. 

From these exact results, we can determine the upper critical dimension of the
roughening transition, at least within the ($2 + \varepsilon$) expansion. In
the directed-polymer representation, the non-equilibrium roughening transition
maps onto a second-order equilibrium transition separating a phase where the
quenched point defects have very little effect on the polymer structure, from a
disorder-dominated (``glassy'') phase. It is generally accepted that for such a
transition the Chayes--Fisher bound \cite{chayes-chayes-fisher-spencer:86} 
$\nu > 2/d$ applies. Assuming that this is valid for the above crossover length
scale as well, and furthermore that there appear no non-perturbative
corrections in higher dimensions, inspection of Eq.~(\ref{crcrex}) tells us
that the correlation length exponent reaches its lower bound $\nu_{\rm min} =
2/d$ at four dimensions, where it then acquires its presumable mean-field value
$\nu_c^{\rm mf} = \frac{1}{2}$. Hence the {\em upper critical dimension for the
roughening transition} is $d_{uc} = 4$.

\subsubsection{Scaling behavior in the smooth phase} 

Let us now discuss the scaling behavior in the smooth phase. We have seen that
for small initial values of $v$ and $u > u_c$, the RG flow leads into the
strong-coupling regime, which has hitherto resisted any successful analytic
approaches. Numerical simulations have found $\chi > 0$, and one therefore
expects this phase to be characterized by a {\em rough} interface. In the
converse situation, $u < u_c$, {\em both} non-linear couplings $u$ and $v$
flow to zero, see Fig.~\ref{mnflow}(b), which renders the scaling relation
(\ref{screlg}) as well as (\ref{screlf}) invalid.  In order to obtain the
correct scaling behavior in the ensuing smooth phase, a more detailed
investigation of the flow of the ratio $w = v / u$ is required.  In terms of
the variables $u$ and $w$, the gamma and beta functions become
\begin{eqnarray}
  \gamma &=& - \frac{u}{2} \, (1 + w)^2 \ ,
\label{mgamma} \\
  \beta_u(u,w) &=& u \left[ \varepsilon - \frac{u}{2}\, (1 + w)^2 \right]\ ,
\label{mbetag} \\
  \beta_w(u,w) &=& w \left[ - 2 \rho + \frac{u}{2} \, (1 + w)^2 \right] \ .
\label{mbetaw}
\end{eqnarray}
Consequently, as $\ell \to 0$ and $u(\ell) \to 0$, $w(\ell) \to \infty$.

\begin{figure}[htb] 
  \hbox{\epsfxsize 0.65\columnwidth \epsfbox{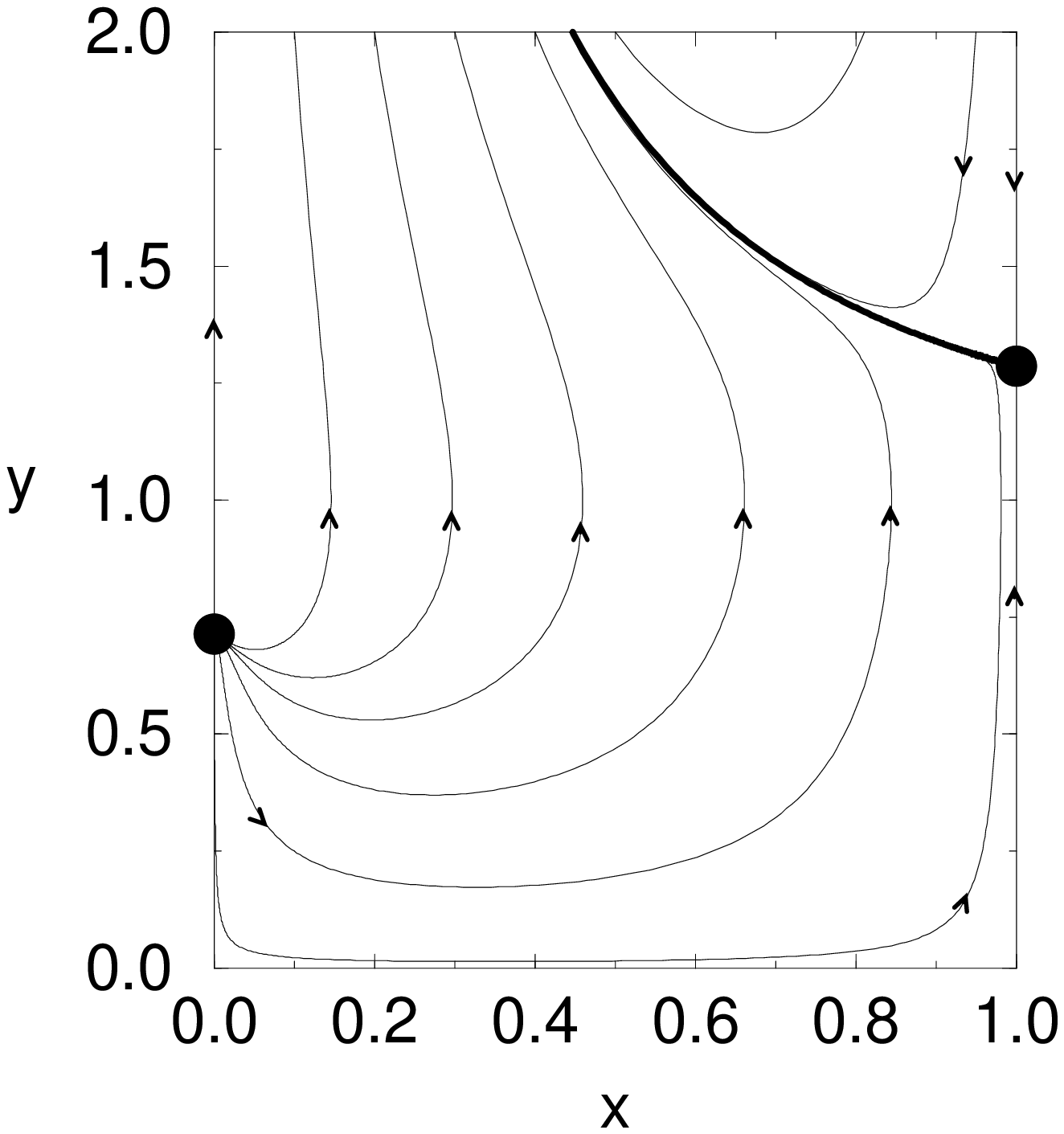} \hspace{-1.5cm}
    \epsfxsize 0.65\columnwidth \epsfbox{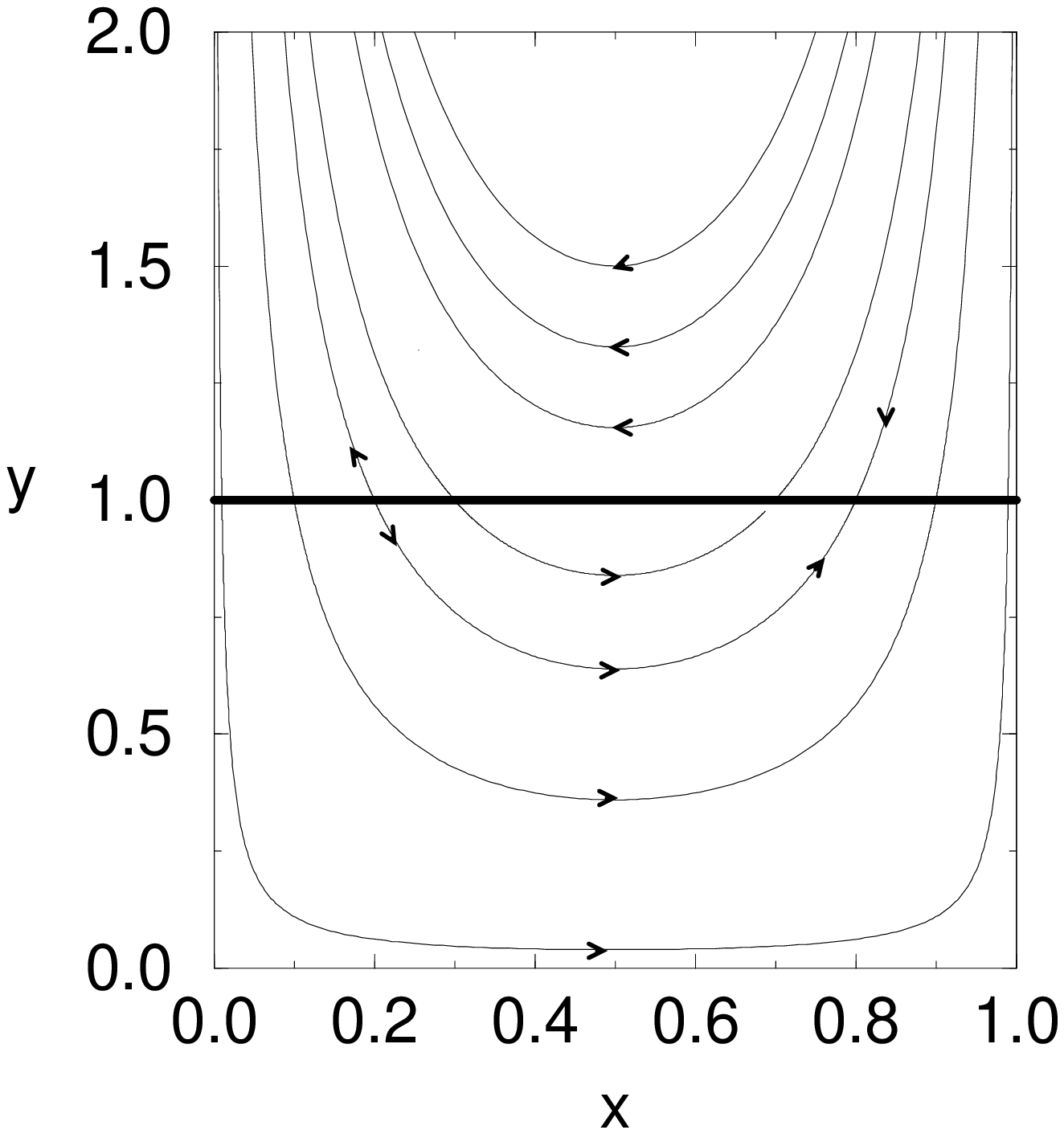}}
  \hbox{\epsfxsize 0.65\columnwidth \epsfbox{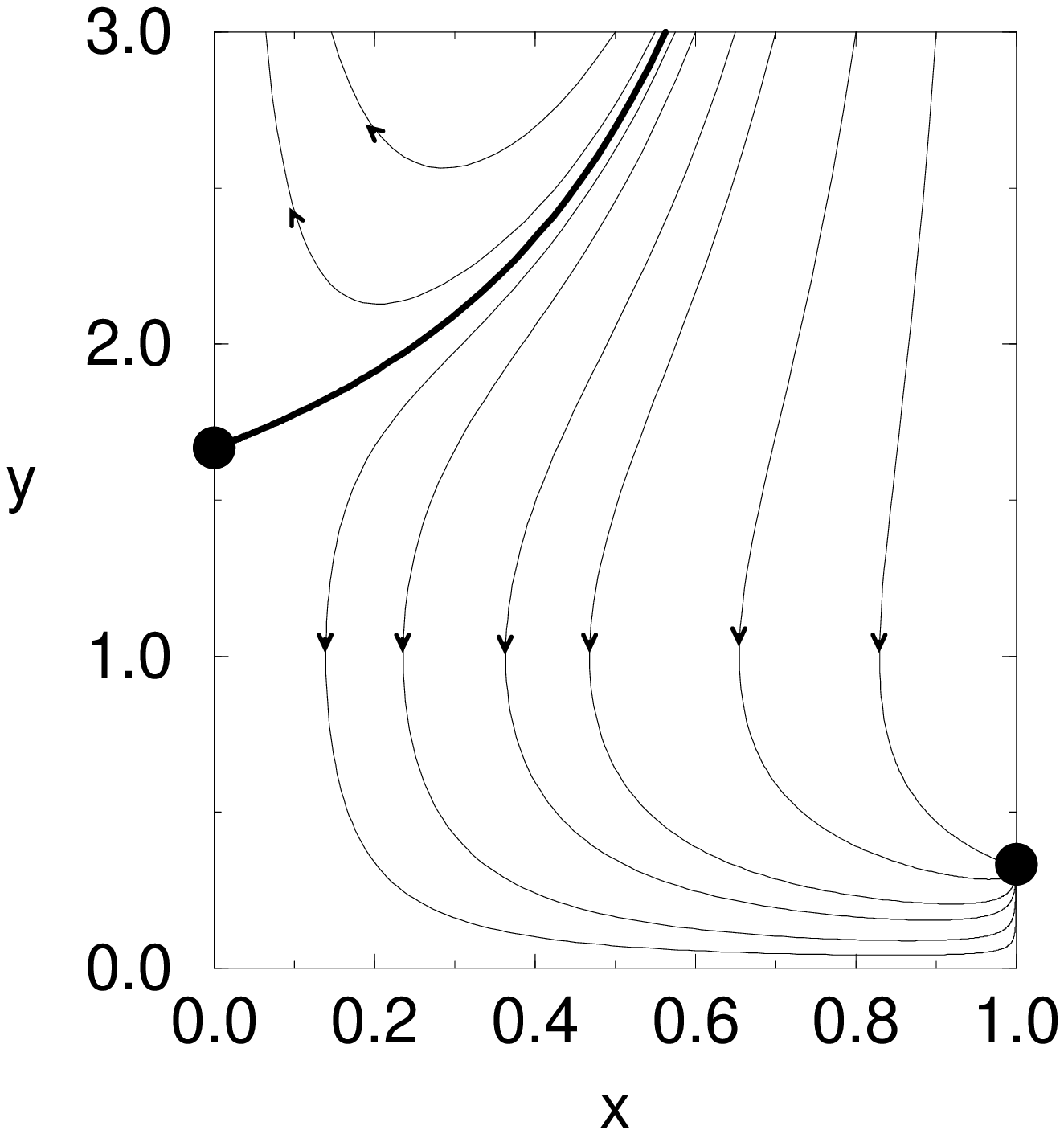} \hspace{-1.5cm}
    \epsfxsize 0.65\columnwidth \epsfbox{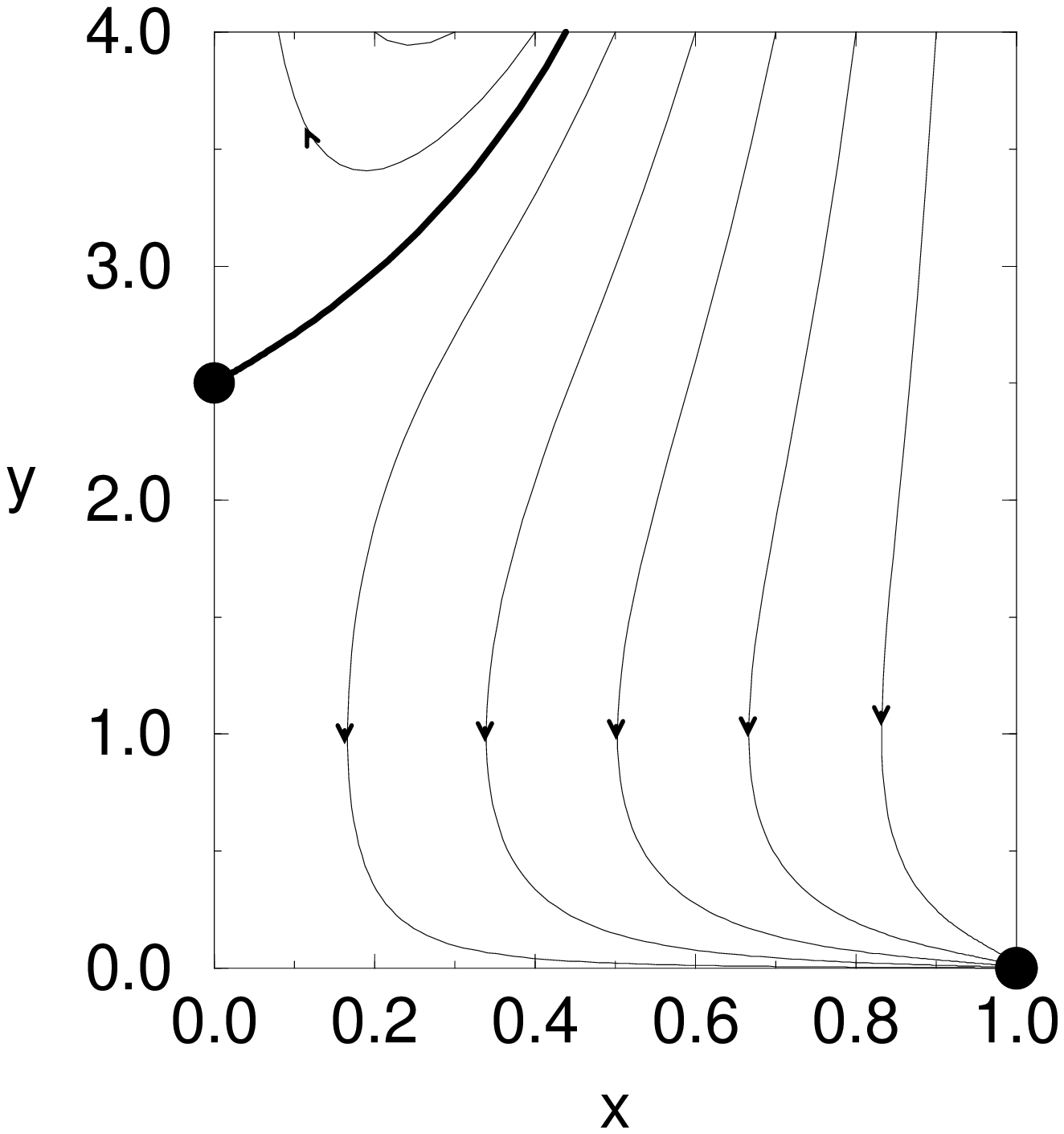}} 
\caption{Flow diagrams in terms of the variables $x$ and $y$ in the regimes
  (a) $d < d_c = 2(1+\rho)$, (b) $d = d_c$, (c) $2(1+\rho) < d < 2(1+2\rho)$, 
  and (d) $d \geq 2(1+2\rho)$.}
\label{xyflow}
\end{figure}

At this point it is important to note that $w$ appeared as a prefactor in the
noise correlation, and therefore multiplies ${\tilde s}^2$ in the action
(\ref{dynfun}). The divergence of $w$ therefore implies that in order to obtain
the correct scaling exponent $\eta$, we need to rescale the fields ${\tilde s}$
and $s$ according to
\begin{equation}
  {\tilde s}' = (1+w)^{\frac{1}{2}} \, {\tilde s} \ , \; {\rm and} \quad
  s' = (1+w)^{-\frac{1}{2}} \, s \ .
\label{fdresc}
\end{equation}
The multiplicative factor $1+w$ then induces an additional term in the relation
between the static exponent $\eta$ and the fixed-point value of $\gamma$,
namely
\begin{equation}
  \gamma' = \gamma + \frac{\beta_w}{1+w} \ , \quad \eta = {\gamma'}^* \ .
\label{scfdrn}
\end{equation}
At the roughening transition, $w \to 0$, and $\eta_c = \gamma^* = -\varepsilon$
is recovered. In the smooth phase, where $w \to \infty$ and $v \to 0$, one 
obtains, however,
\begin{equation}
  \eta_{\rm sm} = - 2 \, \rho \ .
\label{etaexp}
\end{equation}
With $\zeta = 0$ and Eq.~(\ref{rghexp}), we arrive at the scaling exponents in 
the smooth phase
\begin{equation}
  \chi_{\rm sm} = 1 + \rho - \frac{d}{2} \leq 0 \ , \quad z_{\rm sm} = 2 \ .
\label{smoexp}
\end{equation}
Notice that the roughness exponent is negative, which describes a smooth
interface, and furthermore that its actual value is determined by the power-law
correlated noise. When $\rho = (d-2)/2$, the long-range noise is prepared in
such a way that the roughness exponents in the smooth phase and at the 
roughening transition coincide, $\chi_{\rm sm} = \chi_c = 0$. This special
degenerate situation corresponds to $d = d_c$, and therefore the scaling
behavior in the smooth regime cannot be distinguished from the critical
behavior; accordingly, the coupling $v(\ell)$ is constant, see 
Eq.~(\ref{mnbetf}). In the flow diagram, we should thus find a {\em fixed line}
for $d=d_c$, compare Fig.~\ref{xyflow}(b) below.

We shall now analyze the flow equations in more detail, and determine the
correction-to-scaling exponents $\omega_1$ and $\omega_2$.  The latter are
given by the eigenvalues of the stability matrix (\ref{stamat}), evaluated in
the smooth phase, and must both be positive. We shall thus uncover an 
additional subtlety in the correlated-noise KPZ problem, namely the emergence
of {\em two distinct} smooth regimes characterized by different
correction-to-scaling exponents. To this end, we introduce the following new
variables
\begin{equation}
  x = \frac{w}{1 + w} \ , \quad y = \frac{1}{4 \rho} (1 + w)^2 u \ .
\label{newvar}
\end{equation}
The parameter $x$ varies in the interval $[0,1]$ and describes the relative
strength of the short-range and power-law correlated noise contributions. The
variable $y$ basically coincides with the short-range coupling $u$ at the
uncorrelated noise fixed point, where $w = 0$, and furthermore represents the
appropriate combination in the long-range regime. In addition, it is convenient
to re-parametrize the RG flow according to
\begin{equation}
  \ell = e^{- l / 2 \rho} \ , \quad 
  \ell \, \frac{d}{d \ell} = - 2\rho \, \frac{d}{dl} \ .
\label{repflo}
\end{equation}
In terms of this new variable, the RG flow equations for $x$ and $y$ become
\begin{eqnarray}
  \frac{dx}{dl} &=& x \, (1-x) \, (1-y) \ , 
\label{nxflow} \\
  \frac{dy}{dl} &=& y \left[ 1 - \frac{\varepsilon}{2\rho} + 
    (2x-1) \, (1-y) \right] \ ,
\label{nyflow}
\end{eqnarray}
and the stability matrix reads
\begin{eqnarray}
  &&\Omega(x,y) = 
\label{nstmat} \\ 
  &&\qquad \left( \begin{array}{cc} - (2x-1) \, (1-y) & -x \, (1-x) \\
      2 y \, (1-y) & - \frac{\varepsilon}{2 \rho} + 2x - 2 \, (2x-1) \, y \\
    \end{array} \right) \ . \nonumber
\end{eqnarray}

The flow equations (\ref{nxflow}) and (\ref{nyflow}) have four fixed points:
\noindent
(i) $(x,y) = (0,0)$: Edwards--Wilkinson fixed point, with scaling exponents
$\chi = 1 - \frac{d}{2}$, and $z = 2$. The associated eigenvalues of 
$\Omega(x,y)$ are $1$ and $-\varepsilon/2\rho$, and this Gaussian fixed point
is thus {\em unstable} with respect to the long-range noise correlations ($x$
direction) for $d > 2$.

\noindent
(ii) $(x,y) = (0,\varepsilon/2\rho)$: short-range KPZ fixed point, with
stability matrix eigenvalues $1 - \varepsilon / 2\rho$ and $\varepsilon/2\rho$.
This fixed point is {\em unstable} in both directions for $d < d_c$, and 
describes the roughening transition for $d > d_c = 2 (1+\rho)$, with the
critical exponents (\ref{critex}). The crossover and correction-to-scaling
exponents (\ref{crcrex}) and (\ref{crscex}) follow from the above eigenvalues
through multiplication with the factor $-2\rho$, which originates from the
re-parametriza\-tion (\ref{repflo}).

\noindent
(iii) $(x,y) = (1,0)$, with eigenvalues $-1$ and $2 - \varepsilon / 2\rho$:
unstable for $d < 2 (1 + 2\rho)$, stable for $d > 2 (1 + 2\rho)$.
      
\noindent
(iv) $(x,y) = (1,2 - \varepsilon / 2\rho)$, with eigenvalues 
$1 - \varepsilon/ 2\rho$ and $- (2 - \varepsilon / 2\rho)$: unstable for 
$d < d_c$ and $d > 2 (1 + 2\rho)$, stable for $2(1+\rho) < d < 2 (1 + 2\rho)$.

\noindent
The fixed points (iii) and (iv) therefore describe the smooth phase with
scaling exponents (\ref{smoexp}) in the regimes $d > 2 (1 + 2\rho)$ and $d_c =
2(1+\rho) < d < 2 (1 + 2\rho)$, respectively, which are characterized by {\em
  different correction-to-scaling exponents}.

\begin{figure}
  \centerline{\epsfxsize 0.75\columnwidth \epsfbox{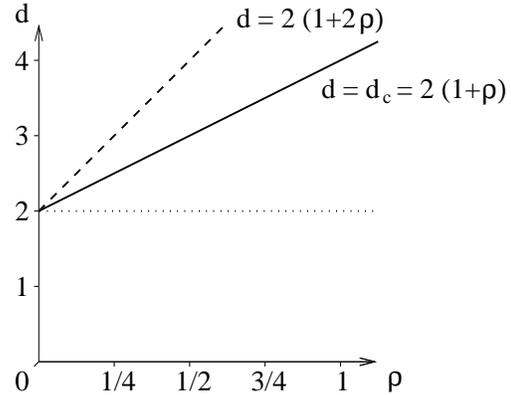}}
\caption{Different regimes for the long-range correlated noise KPZ problem in
  the $(\rho,d)$--plane.}
\label{regime}
\end{figure}

To summarize, we have to distinguish between four distinct regimes, as depicted
in the $(\rho,d)$--plane in Fig.~\ref{regime}.

\noindent
(a) For $d < d_c = 2(1+\rho)$, there exists no non-equilibrium phase transition
from a smooth into a rough phase; the RG flow runs away into the 
strong-coupling regime $y \to \infty$. To what extent this situation is 
accessible via renormalized perturbation theory, is to some extend still an 
open question and will be further discussed in the final section \ref{discs}.

\noindent
(b) Precisely at the critical dimension $d = d_c$, or $\rho = (d-2)/2$, there
appears a {\em fixed line} at $y = 1$, which is unstable for $x < \frac{1}{2}$,
and stable for $x > \frac{1}{2}$. The RG trajectories are hyperbolae 
$y = C / x(1-x)$. The flow separatrix is $x < \frac{1}{2}: y = 1$, 
$x > \frac{1}{2}: y = 1 / 4x(1-x)$.
    
\noindent
Above the critical dimension $d_c$, there appears a non-equilibrium transition
separating a long-range noise dominated smooth phase from a perturbatively
non-accessible strong-coupling rough phase. This roughening transition is 
described by a separatrix in the flow diagram, terminating at the short-range 
KPZ fixed point $(x,y) = (0,\varepsilon/2\rho)$.

\noindent
(c) For $d_c = 2(1+\rho) < d < 2(1+2\rho)$, the stable fixed point 
characterizing the smooth phase is $(x,y) = (1,2 - \varepsilon / 2\rho)$, with
associated correction-to-scaling exponents
\begin{equation}
  \omega_1 = d - 2 (1+\rho) \ , \; {\rm and} \quad 
  \omega_2 = 2 (1+2\rho) - d \ .
\label{corsc1}
\end{equation}

\noindent
(d) For $d \geq 2(1+2\rho)$, the stable fixed point governing the smooth
phase becomes $(x,y) = (1,0)$, characterized by
\begin{equation}
  \omega_1 = d - 2 (1+2\rho) \ , \; {\rm and} \quad \omega_2 = 2 \rho \ .
\label{corsc2}
\end{equation}

The strong-coupling phase with presumably positive roughness exponent cannot be
addressed with our methods. However, the numerical solutions of the flow
equations would suggest that $x \to 0$ in the rough phase, and hence that the
long-range noise correlations are {\em irrelevant} in the strong-coupling
regime. Yet, of course we cannot exclude either that there might exist a 
further crossover line separating this short-range regime from another scaling
region, where the initial long-range noise correlations dominate.

\section{General spatial noise correlations}
\label{sec:bseqn}

We can readily generalize our analysis of the KPZ equation (\ref{kpzeqn}) to
{\em arbitrary spatial noise} $R_0 ( \vec{x}-\vec{x}' )$, provided the 
correlations of the stochastic forces remain {\em local in time}, where from 
now on we indicate the bare form of the noise correlations by an index ``0''.
As of course the structure of the ensuing field theory has not changed as 
compared to Sec.~\ref{sec:rough}, there still is no propagator renormalization,
and therefore $z = 2$. Upon rescaling the vertex function 
$\Gamma_{22}(\vec{k},\vec{k}'; \mu^2)$ by a factor $g^2/D$, the Bethe--Salpeter
equation for the ``renormalized'' noise reads
\begin{eqnarray}
  &&R (\vec{k},\vec{k}'; \mu^2) = 
    R_0(\vec{k}-\vec{k}') \nonumber \\
  &&\qquad + \frac{g^2}{4} \int \! \frac{d^dp}{(2\pi)^d} \, 
  \frac{R_0(\vec{k}-\vec{p})}{p^2 + \mu^2} \, 
  R(\vec{p},\vec{k}'; \mu^2) \ .
\label{inteqn}
\end{eqnarray}
Note that this {\em exact} equation allows us to determine the full noise
correlation function $R (\vec{k},\vec{k}'; \mu^2)$ for any given bare noise
$R_0 (\vec{k})$. For simplicity, in the following we shall restrict ourselves 
to the case where $\vec{k}'=0$.

\subsection{Mathematical structure of the Bethe-Salpeter equation}

Upon defining as $f(k;\mu^2) = R (\vec{k},\vec{k}'=\vec{0};\mu^2) / R_0 (k)$ 
the ratio of the ``renormalized'' and the bare noise correlator, and carrying 
out the angular integrations, the Bethe--Salpeter equation assumes the form
\begin{eqnarray}
  f(k;\mu^2) = 1 + \lambda \int_0^\Lambda \! dp \, K(k,p;\mu^2) \, f(p;\mu^2) 
  \ ,
\end{eqnarray}
where we have introduced an upper momentum cutoff $\Lambda$, and furthermore 
defined the integral kernel
\begin{eqnarray}
 K(k,p;\mu^2) = \frac{p^{d-1} R_0 (p)}{(p^2 + \mu^2) R_0 (k)} \, 
  \int \! d \eta  \,  R_0 ( |\vec{k} - \vec{p}|)  \ , 
\end{eqnarray}
with $\eta$ denoting the cosine of the angle between the wave vectors $\vec{k}$
and $\vec{p}$.

Now the Bethe-Salpeter equation has the structure of a {\em Fredholm integral
equation of the second kind}. The parameter $\lambda$ in the integral equation,
\begin{equation}
 \lambda = \frac{S_d}{8} \, g^2 \ ,
\end{equation}
is proportional to the effective interaction parameter of the KPZ equation, and
the kernel $K(k,p;\mu^2)$ is a functional of the bare noise correlator $R_0(k)$
and completely determined by it. In general, the existence of a solution for a 
Fredholm integral equation depends on the strength of the interaction parameter
$\lambda$. It is known that a perturbative solution in terms of a Neumann 
series exists for sufficiently small $\lambda$, provided the kernel 
$K(k,p;\mu^2)$ is bounded. Note that the von-Neumann series is nothing but the
perturbation expansion symbolically depicted in Fig.~\ref{betsal}. The critical
value $\lambda_c$, where a perturbative solution breaks down, is given by the
eigenvalue of the homogeneous equation $f(k;\mu^2=0) = \lambda \int_0^\Lambda
dp \, K(k,p;\mu^2=0) f(p;\mu^2=0)$. Physically, this eigenvalue corresponds to 
the critical parameter value of the effective coupling constant where a phase
transition from the smooth to the rough phase occurs. Given a certain 
functional form of the bare noise correlator, this allows for an {\em exact}
determination of the location of the roughening transition, i.e., the structure
of the phase diagram. In this way, we can lucidly relate the mathematical
properties of the Bethe--Salpeter or Fredholm integral equation to the 
physical structure emerging by means of the renormalization-group analysis.

Formally, there are also solutions of the Fredholm integral equation above the
critical value of the $\lambda$--parameter which cannot be obtained in the form
of a von-Neumann series (perturbation expansion). From the numerical solutions 
(see below), it turns out that the noise correlations then become negative in a
certain wave-vector regime. The physical significance of these regions of 
negative correlations is unclear, and it would be mere speculation to assign
any meaning to them associated with the strong-coupling phase.

\subsection{Numerical solution of the Bethe-Salpeter equation}

In this section, we study the numerical solution of the Bethe--Salpeter 
integral equation for $\mu^2=0$ and two different forms of the wave-vector 
dependence for the bare noise correlation function, namely (i) Gaussian, and
(ii) power-law correlated noise.

\subsubsection{Gaussian noise} 

\begin{figure}[htb]
  \centerline{\epsfxsize 0.6\columnwidth \epsfbox{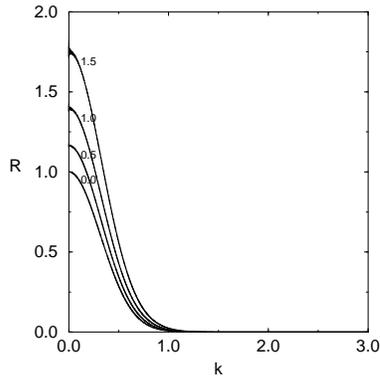}}  
\caption{Renormalized noise correlator $R(k) = f(k) R_0 (k)$ calculated from 
  the Bethe--Salpeter equation in $d=3$ with a bare Gaussian noise 
  $R_0(k) = \exp (-5 k^2)$ for values $\lambda = 0$, $0.5$, $1.0$ and $1.5$ of
  the interaction parameter (as indicated in the graph). The cutoff was fixed 
  to $\Lambda = 3$.}
\label{fig:gauss_noise}
\end{figure}

We start with a bare noise correlation function given by a Gaussian
\begin{eqnarray}
  R_0(k) = \exp \left( - \alpha k^2 \right) \ .
\end{eqnarray}
Then the integral kernel takes the form
\begin{eqnarray}
  K(k,p) = \frac{p^{d-4}}{2 k} e^{-2 \alpha p^2} \sinh (2 \alpha k p) \ .
\end{eqnarray}
The ensuing integral equation may be solved by means of standard numerical 
procedures, leading to the results for the renormalized noise correlator 
$R (k) = f(k) R_0 (k)$ shown in Fig.~\ref{fig:gauss_noise} in three dimensions,
and with noise parameters $\alpha = 5$, $\Lambda = 3$ and interaction 
parameters $\lambda = 0$, $0.5$, $1.0$ and $1.5$, respectively. It is seen that
upon increasing $\lambda$ towards ist critical value of $\lambda_c \approx 
1.515$, the amplitude of the renormalized noise at wave vector $k=0$ diverges 
as a power law with an exponent given by approximately $-1$.

\subsubsection{Power law noise} 

\begin{figure}[htb]
  \centerline{\epsfxsize 0.6\columnwidth \epsfbox{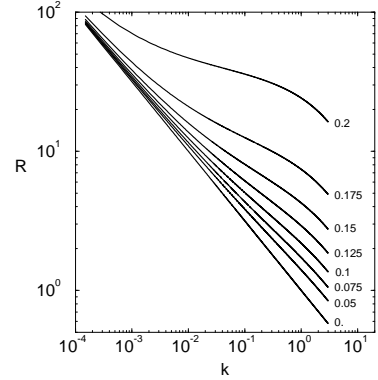}}
  \caption{Renormalized noise correlator $R (k) = f(k) R_0 (k)$ calculated from
    the Bethe--Salpeter equation in $d=3$ with a power law noise 
    $R_0(k) = k^{-\frac{1}{2}}$ for values $\lambda = 0$, $0.05$, $0.075$, 
    $0.1$, $0.125$, $0.15$, $0.175$ and $0.2$ of the interaction parameter 
    (as indicated in the graph). The cutoff was fixed to $\Lambda = 3$.}
\label{fig:power_noise}
\end{figure}

Next we consider the solution of the Bethe--Salpeter equation starting from a
bare power law noise of the form
\begin{eqnarray}
  R_0 (k) = k^{-2 \rho} \ .
\end{eqnarray}
Then the integral kernel becomes
\begin{eqnarray}
  K(k,p) = \frac{k^{2 \rho-1} p^{d-4-2\rho} }{2 (1-\rho)} 
  \left[ (k+p)^{2(1-\rho)} - (k-p)^{2(1-\rho)} \right] \ .
  \nonumber
\end{eqnarray}
With increasing interaction parameter, the spectrum of the noise correlator
shows an overall increase, but at the same time it develops more weight at wave
vectors close to the cutoff $\Lambda$. In Fig.~\ref{fig:power_noise}, a 
solution of the Bethe--Salpeter equation is shown for the parameter values 
$\rho = \frac{1}{4}$ and $\Lambda = 3$ in $d=3$. Again, there is a critical
value $\lambda_c$ where the noise amplitude shows a power law divergence.

The two particular forms of the bare noise correlation functions discussed
above may exemplify the potential of the Bethe--Salpeter equation in 
determining the structure of the phase diagram, and specifically, the line in
parameter space where the roughening transition occurs. If required (e.g., for 
a particular numerical simulation) precise values could be determined allowing 
for quite stringent tests on the theoretical concepts presented in this paper.

\section{Summary and discussion}
\label{discs}

\begin{figure}[htb]
  \centerline{\epsfxsize 0.6\columnwidth \epsfbox{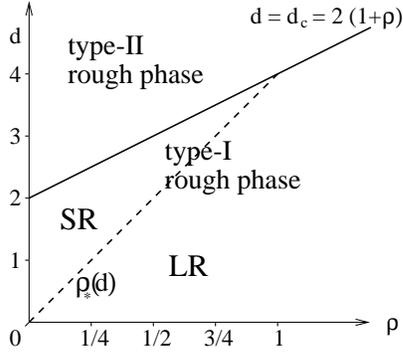}} 
  \vspace{0.5cm}
\caption{``Phase diagram'' for the KPZ equation with correlated noise in the 
  $(\rho,d)$ plane. The solid line, $d=d_c = 2(1 + \rho)$, indicates the 
  {\em exactly} determined boundary line above which the 
  Kardar\--Parisi\--Zhang\--equation with correlated noise shows a 
  {\em non-equilibrium phase transition}, and below which there exists only a 
  {\em rough phase}.}
\label{fig:phase_diagram}
\end{figure}

In this paper, we have studied the Kardar--Parisi--Zhang equation with 
long-range correlated noise (\ref{noidef}). Our results are best summarized in 
the following ``phase diagram'' (Fig.~\ref{fig:phase_diagram}) in the plane 
spanned by the spatial dimension $d$ and the correlation exponent $\rho$ of the
noise. Here, the solid line defined by $d=d_c = 2(1 + \rho)$ indicates the 
{\em exactly} determined boundary line above which the 
Kardar\---Parisi\---Zhang equation with correlated noise shows a 
{\em non-equilibrium roughening transition}, separating a Gaussian (EW) smooth 
regime from a strong-coupling rough phase, and below which there exists only a 
{\em rough phase}.

\subsection{Exact results for the roughening transition above the lower
  critical dimension}

We start our summary with a discussion of the results above the lower critical
dimension $d_c$. Here, there are two stable and one unstable (critical) fixed 
points, describing a smooth and rough phase, respectively, as well as the 
non-equilibrium separating these. Using a stochastic Cole--Hopf transformation,
we were able to derive the following {\em exact results} for the roughening 
transition and the scaling properties in the smooth phase:
\begin{itemize}
\item[a.]  The roughening transition is characterized by an unstable fixed
  point much in the same way as the tempe\-rature-driven phase transition in 
  the non-linear $\sigma$-mod\-el. As a result, which is {\em exact} to all 
  orders in a $(2+\varepsilon)$ expansion, we obtain for the dynamic exponent 
  $z_c = 2$, the roughness exponent $\chi_c = 0$, the crossover exponent 
  $\phi_c = d-2$, and the correction-to-scaling exponent 
  $\omega_c = d - 2(1+\rho)$. While $\phi_c$ describes the divergence of the 
  crossover length scale, $\omega_c$ governs the next-to-leading terms in the 
  corresponding scaling functions.
\item[b.] For effective coupling constants {\em less} than the values 
  corresponding to the unstable fixed point, the flow tends towards a stable 
  fixed point that describes the smooth phase. We find that the long-range 
  fixed point is stable, and that the smooth phase is described by the scaling 
  exponents $z_{\rm sm} = 2$, $\chi_{\rm sm} = 1+\rho-d/2$, and the 
  correction-to-scaling exponents $\omega_1 = d-2(1+\rho)$, 
  $\omega_2 = 2(1+2\rho)-d$ for $2(1+\rho) < d < 2(1+2\rho)$, and 
  $\omega_1 = d-2(1+2\rho)$, $\omega_2 = 2\rho$ for $d \geq 2(1+2\rho)$, 
  respectively.
\item[c.] The flow of the RG trajectories starting with effective couplings
  {\em larger} than the critical value, tends to infinity and seems to converge
  towards short-range noise behavior. Although we cannot exclude that the flow
  returns to effectively long-range noise behavior at larger couplings, this 
  result seems to indicate that long-range noise does not affect the critical 
  behavior of the strong-coupling rough phase above $d_c$.
\item[d.] For {\em any} given initial (``bare'') noise correlator that is local
  in time, the noise correlation function can be determined {\em exactly} using
  a Bethe--Salpeter integral equation which relates the bare to the full noise 
  correlation. The mathematical structure of this Fredholm integral equation 
  reflects the renormalization group results.
\end{itemize}
The rough phase above $d_c$ does not seem to be accessible by any 
perturbational means, and is presumably qualitatively different from the rough
phase below $d_c$, which we are going to discuss next.

\subsection{Exact results and some conjectures on the rough phase below the
  lower critical dimension}

In dimensions $d < d_c = 2 (1+\rho)$, the interface scaling behavior is
governed by a strong-coupling fixed point. Contrary to the case above $d_c$,
there actually do exist some perturbational techniques which allow for the
assessment of the ensuing scaling behavior:
\begin{itemize}
\item[a.] First of all --- as was already noted in 
  Ref.~\cite{medina-hwa-kardar-zhang:89} --- the general structure of the RG
  theory tells us that there are two possible fixed points, one corresponding
  to short-range noise, the other one to long-range noise. In addition, if the
  long-range noise fixed point becomes stable, the critical exponents here can 
  be read off from the RG equation {\em exactly}, without knowing the actual 
  value of the fixed point. One obtains
\begin{equation}
  \chi_{\rm lr} = \frac{2-d+2\rho}{3} \; , \ {\rm and} \quad
  z_{\rm lr} = \frac{4+d-2\rho}{3} \ .
\end{equation}
\item[b.] One way to approach the strong-coupling behavior below $d_c$ is by
  means of a non-minimal renormalization scheme. In one dimension, we establish
  the {\em exact} result that the usual KPZ short-range noise fixed point with
  scaling exponents $\chi_{\rm sr} = \frac{1}{2}$ and $z_{\rm sr} = 
  \frac{3}{2}$ remains stable, provided $\rho < \frac{1}{4}$. For $\frac{1}{4}
  \leq \rho < 1$, we find $\chi_{\rm lr} = (2-d+2\rho)/3$ and $z_{\rm lr} =
  (4+d-2\rho)/3$. These results may also be re-derived within a proper
  $\varepsilon$ expansion using a mapping to driven diffusive systems
  \cite{janssen-schmittmann:98}.
\item[c.] In the $(\rho,d)$--plane, the point $(1,4)$ is a special point since
  it describes the Burgers equation with a non-con\-served noise, i.e., model B
  in the pioneering work by Forster {\em et\/ al.}
  \cite{forster-nelson-stephen:77}. As we have already noted previously in this
  paper and as is explicitly shown in Ref.~\cite{forster-nelson-stephen:77},  
  the infrared behavior of the {\em non-conserved Burgers equation} is governed
  by an infrared-stable (long-range noise) fixed point which can be explicitly
  obtained within a one-loop perturbative calculation around the upper critical
  dimension $d_{\text{uc}}=4$ of this model. Actually, the dynamic exponent 
  obtained via the $\varepsilon$ expansion turns out to be the same as the one 
  resulting from exact arguments (see item a. above), namely 
  $z_{\rm lr}=(2+d)/3$.
\end{itemize}

In the remainder of this section, we discuss the boundary line between the
short-range and long-range noise fixed points {\em below} the lower critical
dimension $d_c$. In order to determine the location of this boundary line, we 
only need to return to the above results, and combine them with established 
facts from the literature. First, we know from the exact solution of the KPZ 
equation in $d=1$ that the critical value of $\rho$ above which the short-range
noise fixed point becomes unstable is $\rho=\frac{1}{4}$. Next, we may ask 
where the boundary between the short-range and long-range fixed points in the
$(\rho,d)$--plane crosses the line marking the lower critical dimension, 
$d_c = 2 (1+\rho)$, compare Fig.~\ref{fig:phase_diagram}. It is actually quite
easy to see that this crossing point is given by $(\rho,d) = (1,4)$, simply by 
noting that for $\rho = 1$, we arrive at the non-conserved Burgers equation 
with associated upper critical dimension $d_c = 4$. In addition, upon adding
noise with $\rho > 1$ to the non-conserved noise of the Burgers equation, i.e.,
at the long-range noise fixed point, the dynamic exponent becomes 
$z = (4 + d - 2 \rho)/3$. Hence it is quite likely that at $d=4$, and for
$\rho \to 1$, this fixed point coincides with the unstable fixed point 
describing the {\em roughening transition} of the KPZ equation, with $z_c = 2$
and $\phi_c = \nu_c^{-1} = 2$ --- in fact, all of these fixed points become
zero at this very particular point in the $(\rho,d)$--plane. Notice that this
essentially excludes the other possible guess, namely that the long-range
noise fixed point from the regime below $d_c$ turns into the 
{\em strong-coupling} fixed point above the roughening transition at 
$(\rho,d) = (1,4)$. Of course, one might still argue that this strong-coupling
fixed point vanishes as well at this point, but this would appear rather
strange.

Having established that the separatrix between the long-range and short-range
noise fixed points has to go through two points, $(\rho,d) = (\frac{1}{4},1)$
and $(1,4)$, we would also like to know the location of this curve in between. 
As a crude approximation to the actual functional form of the separatrix 
$\rho_*(d)$, one may take a linear interpolation between the two exactly 
determined points, i.e., $\rho_*(d) = d/4$. In fact, recent computer 
simulations \cite{li:97} found $\rho_*(d=2) \approx 0.5$, which confirms that
this simple interpolation formula may already be quite close to the actual 
function $\rho_*(d)$. Note that the knowledge of the separatrix enables us to 
determine the critical exponents of the short-range noise strong-coupling fixed
point, if one makes the quite plausible assumption that the critical exponents 
are continuous functions of $\rho$. Inserting the linear interpolation formula 
into the exact value for the long-range dynamic exponent, this would give
\begin{eqnarray}
  z = \frac{4+d-2(d/4)}{3} = \frac{8+d}{6} \ .
\end{eqnarray}
Interestingly, this simple estimate turns out be {\em identical} with the
result obtained from a functional renormalization group analysis by
Halpin-Healy \cite{halpin-healy:90}, and a more recent mode-coupling study by
Bhattacharjee \cite{bhattacharjee:98}; and of course, it reproduces the exact
one-dimensional value $z = 3/2$.

What can we learn from this? If we assume that the critical exponents are 
really continuous over the whole parameter range of $\rho$ and $d$, this would 
imply that we would actually know the values of the critical exponents in the 
rough phase, both below and above the lower critical dimension $d_c$. As an 
immediate consequence, this would tell us that $4$ is the upper critical 
dimension for both the rough phase and the roughening transition in accordance 
with some recent speculations \cite{laessig-kinzelbach:97}. However, because
of the divergence of the strong-coupling fixed point in the short-range KPZ
equation at $d_c = 2$, which indicates {\em non-analytic} behavior at the lower
critical dimension \cite{frey-taeuber:94,laessig:95,taeuber-frey:95}, we do 
{\em not} think that this assumption is correct. Instead, we suggest the 
following scenario:

\begin{description}
\item[a.] One has to distinguish between two fundamentally different notions of
  rough phases, one {\em below} the lower critical dimension 
  $d_c = 2 ( 1 + \rho)$, and the other one {\em above}; in the following, we 
  shall denote these two distinct rough phases as type--I and type--II, 
  respectively.
\item[b.] The dashed line in Fig.~\ref{fig:phase_diagram} marks the 
  {\em separatrix} between the short-range and long-range noise fixed point of 
  the type--I rough phase {\em below} the lower critical dimension. The 
  critical exponents are continuous at this line. 
\item[c.] Anticipating that the separatrix between long-range and short-range
  noise fixed points is approximately given by $\rho = d/4$, the scaling
  exponents in the sliver below the solid line marking the lower critical 
  dimension $d_c(\rho)$ of the problem become
  \begin{equation}
    z = \frac{d+8}{6} \; , \quad {\rm and} \quad \chi = 2 - z \ .
  \end{equation}
\item[d.] It is possible to assess the scaling behavior in this type--I regime 
  in various ways. First of all, exact relations tell us that once we are below
  the dashed line, i.e., asymptotically at the {\em long-range} fixed point, 
  all exponents are known exactly \cite{fnote3}. In addition, a perturbative
  expansion may be performed for the non-conserved Burgers equation near the
  upper critical dimension $4$ of this model. In the regime where the
  {\em short-range} noise fixed point becomes stable, several approaches have 
  turned out to be useful: (i) a fixed dimension RG approach outlined in 
  section \ref{1dren}, see also Ref.~\cite{frey-taeuber:94}; (ii) mode-oupling 
  theory in $d=1$ \cite{hwa-frey:91,frey-taeuber-hwa:96}; and (iii) a 
  perturbative mode-coupling approach near $d=4$ \cite{bhattacharjee:98}. 
  Furthermore, it seems that the functional RG approach by Halpin-Healy 
  \cite{halpin-healy:90} also applies in this regime, since it yields the 
  identical result for the dashed line as our linear interpolation between two 
  exactly known results. It would be interesting to better understand the
  apparent relationship between the mode-coupling and functional RG approaches,
  which is suggested by the above results. It is important to note at this
  point that whenever one allows, even implicitly, for long-range correlated 
  noise in the KPZ problem, the lower critical dimension is shifted 
  {\em upwards}. One must then be very careful {\em not} to mistake the ensuing
  type--I strong-coupling scaling behavior with the type--II strong--coupling
  phase above the roughening transition.  
\item[e.] As an immediate consequence of the discussion in item d., the upper
  critical dimension of the type--I rough phase {\em below} the lower critical 
  dimension is $4$.
\item[f.] The scaling exponents of the rough phase are {\em not} continuous at
  the lower critical dimension. As indicated by the divergence of the effective
  coupling constant within a fixed-dimension 2-loop calculation 
  \cite{frey-taeuber:94}, there seems to exist a qualitatively new 
  strong-coupling behavior once a roughening transition between a smooth and a
  rough phase appears. The behavior of the type--II rough phase {\em above} the
  lower critical dimension ist most likely not even described in terms of the 
  continuous KPZ model, and the detailed scaling properties might well be
  {\em non-universal}. Recent investigations, where various non-universal 
  aspects of the type--II strong-coupling phase of the short-range KPZ equation
  are observed 
  \cite{drossel:96,marsili-bray:96,newman-bray:96,newman-swift:97}, tend to 
  support this view. 
\item[g.] The critical exponent at the roughening transition, as well as the
  noise correlations, can be obtained exactly by means of a stochastic 
  Cole--Hopf transformation leading to an exact Bethe--Salpeter integral 
  equation for the noise correlation function. The upper critical dimension of
  the roughening transition is $4$.
\end{description}

In summary, we suggest that the above scenario provides a coherent picture for 
most of the available numerical and analytical results on the KPZ problem. In 
particular, it resolves the long-standing discrepancies between the numerical 
observation that there is no indication of an upper critical dimension near 
four dimensions, and several recent speculations that despite this fact $d=4$ 
should be the upper critical dimension of the rough phase. We shall not attempt
to judge on the validity of several recent claims in favor of the upper 
critical dimension $4$. Our view presented above would imply that all of these 
investigations are mainly concerned with the behavior of the {\em type--I} 
rough phase {\em below} the lower critical dimension $d_c$, shifted upwards as
a consequence of the appearance of effectively long-range power-law noise
correlations. For the mode-coupling approaches it is clear how this comes
about. Here, one starts by construction with long-range correlated noise, and 
subsequently tries to find a self-consistent solution. This is likely to almost
automatically restrain the mode-coupling approach to the regime below the line 
marking the lower critical dimension. Hence these methods rather describe the 
type--I strong-coupling phase rather than the type--II strong-coupling phase 
they were aiming at.

To our opinion, besides the precise location of the separatrix $\rho_*(d)$ 
below the lower critical dimension $d_c$, the truly open issue in the 
non-equilibrium surface growth problem described by the KPZ equation, thus
constitutes the scaling behavior in the {\em type--II} strong-coupling phase 
beyond the non-equilibrium roughening transition that appears for $d > d_c$. 
Future work, probably based on the original discrete growth models, should 
address the fundamental questions whether, and in which sense there is 
universality in this regime, and furthermore whether the continous KPZ equation
does in fact provide an adequate continuous representation of the ensuing
physics.

\begin{acknowledgement}
  We benefited from discussions with J.\ Cardy, T.\ Halpin-Healy, and 
  K.\ Wiese.
  H.K.J.\ acknowledges support from the SFB 237 (``Unordnung und gro\ss e
  Fluktuationen'') from the Deutsche For\-schungs\-gemein\-schaft (DFG).
  U.C.T.\ and E.F.\ acknowledge support from the Deutsche
  Forschungsgemeinschaft through a habilitation and a Heisenberg fellowship,
  respectively, DFG-Gz. Ta 177 / 2-1,2 and Fr 850 / 3-1.
\end{acknowledgement}


\appendix

\section{Detailed balance conditions in terms of the Fokker--Planck equation}
\label{appa}

The validity of detailed balance in $d=1$ dimensions can also be derived in the
Burgers representation within the framework of the associated Fokker--Planck 
equation. The Burgers equation reads
\begin{equation}
  \frac{\partial \vec{u}}{\partial t} + 
  \frac{D g}{2} \vec{\nabla} \vec{u}^2 = 
  D \vec{\nabla}^2 \vec{u} + \vec{\eta} \ ,
\label{burgeq}
\end{equation}
where $\vec{u} = - \vec{\nabla} s$, and for $w = 0$
\begin{equation}
  \langle \eta_\alpha (\vec{x},t) \eta_\beta (\vec{x}',t') \rangle = 
  - 2 D \nabla_\alpha \nabla_\beta \delta(\vec{x}-\vec{x}') \delta(t-t') \ .
\end{equation}
For $D g = 1$, the Langevin equation (\ref{burgeq}) reduces to the noisy
Burgers equation. For $g = 0$, the stationary distribution for the diffusive
field $\vec{u}$ with short-range conserved noise is obviously
given by
\begin{equation}
  {\cal P}_{st} [\vec{u}] \propto 
  \exp \left[ - \frac{1}{2} \int \vec{u} (\vec{x})^2 d^dx \right] \ .
\label{statpd}
\end{equation} 
By writing down the Fokker--Planck equation associated with the stochastic
process (\ref{burgeq}), one readily derives the following condition that the
stationary distribution (\ref{statpd}) remain unaltered by the non-linear term
$\propto g$ \cite{deker-haake:75},
\begin{eqnarray}
  0 &=& \int d^dx \ \frac{\delta}{\delta u(\vec{x},t)} 
    \left( \vec{\nabla} \vec{u}^2 (\vec{x},t) 
    {\cal P}[\vec{u}] \right) \nonumber \\
  &=& \int d^dx \ \left( 2 \vec{\nabla} \cdot \vec{u} - 
                         \vec{u} \cdot \vec{\nabla} \vec{u}^2 \right) 
  {\cal P}[\vec{u}] \ .
\label{intcon}
\end{eqnarray}
While the first term in this expression vanishes, provided appropriate boundary
conditions are employed, the second contribution is generally non-zero in any
substrate dimension $d \not= 1$. In one dimension, however, it reduces to 
$\int (d u^3 / dx) dx = 0$, and (\ref{statpd}) is indeed the stationary 
distribution even for the non-linear problem.

\section{Consequences of Galilean Invariance}
\label{appb}

In this appendix, we will exploit the invariance of the dynamic functional
${\cal J}$ in the Cole--Hopf representation under a Galilean transformation.  
The transformation of the fields introduced in (\ref{holkfs}), (\ref{holkts}) 
and (\ref{eq:galilean}) reads to linear order in the parameter $\vec{v}$,
\begin{eqnarray}
  n\left( \vec{x},t\right) &\rightarrow &n\left( {\bf x+}2D\vec{v}t,t\right)
  \exp \left( {\bf v\cdot x}\right)  \nonumber \\
  \tilde{n}\left( \vec{x},t\right) &\rightarrow &\tilde{n}\left( {\bf x+}2D%
  \vec{v}t,t\right) \exp \left( -{\bf v\cdot x}\right) \ . \label{a}
\end{eqnarray}
Thus the Galilei transformation is represented by a type of gauge 
transformation with a gauge function linear in $\vec{x}$. Including 
translational invariance, it follows that each invariant term of ${\cal J}$
consists of an equal number of $n$ and $\tilde{n}$ fields, because the gauge
factors must completely canceled. Therefore, the dynamic functional can be 
written in the form 
\begin{equation}
  {\cal J}=\int dt\int d^{d}x\,\left\{ \tilde{n}\dot{n}+D \vec{\nabla} 
  \tilde{n} \nabla n{\bf -}D{\cal V}\left[ \tilde{n}n\right] \right\} \ .  
\label{b}
\end{equation}
In a fully local theory, we can expand with respect to powers of the fields and
their derivatives,
\begin{equation}
  {\cal V}\left[ \tilde{n}n\right] = g_{2}\left( \tilde{n}n\right)^{2}+
  g_{3}\left( \tilde{n}n\right) ^{3}+\cdots +g_{2}^{\prime }\left( \vec{\nabla}
  \left( \tilde{n}n\right) \right) ^{2}+\cdots \ . \label{c}
\end{equation}
Below and near $d_c={2}$, only the first coupling constant $g_{2}=g^{2}$
acquires positive scaling dimension. As usual, this demonstrates the
renormalizability of the theory with only $g_{2}\neq 0$, i.e., the original
Burgers--KPZ theory, or, in other words, the universality of this theory with
respect to a representation of Galilei invariance through a stochastic field
theory with one scalar field and its conjugate response field.

For $d > d_c=2$, a renormalized perturbation theory can be constructed in
analogy with the non-linear $\sigma$-model by means of an $\varepsilon$
expansion with $\varepsilon =d-2>0$. Here, the perturbation theory for the 
vertex function $\Gamma _{2,2}$ is completely resumable with the aid of the 
Bethe--Salpeter equation, and leads to {\em exact} renormalizations. However, 
the negative scaling dimension of $g_{2}$ for $\varepsilon >0$, as well as the
emergence of higher-order invariants in the expansion of ${\cal V}$ in 
Eq.~(\ref{c}), remind us that such a construction is in generally only possible
if one lets $\varepsilon \rightarrow 0$ {\em first}, and only afterwards the 
order of the perturbation expansion $N\rightarrow \infty$. This is true because
at order $N=O\left( 2/\varepsilon \right) $, new primitive divergences are 
generated which have to be eliminated by counterterms taking the functional 
form of the higher-order contributions. Luckily, in the case of the 
Burgers--KPZ field theory in the Cole--Hopf representation in the smooth phase,
i.e., below the roughening transition, the perturbation theory of the 
renormalization factors stops already at second order. But it is generally
dangerous, for example, to speculate on $d=4$ as a new higher critical 
dimension by means of purely perturbational arguments only.

Let us now consider the general form of an invariant bi-quadratic term in the
dynamic functional ${\cal J}$ or in the generating functional $\Gamma$ for the 
vertex functions,
\begin{equation}
  {\cal K}=\int_{x,t} \! \! \tilde{n}\left( \vec{x}_{1},t_{1}\right) 
  \tilde{n}\left( \vec{x}_{2},t_{2}\right) K\left( \left\{ \vec{x},t\right\}
  \right) n\left( \vec{x}_{3},t_{3}\right) n\left( \vec{x}_{4},t_{4}\right) 
  \ , \nonumber
\end{equation}
like the correlated-noise term in ${\cal J}$, i.e.,  the contribution 
$\Gamma_{2,2}$ to $\Gamma$ in our model. The coupling function $K$ must obey
the Galilean invariance condition (to linear order in the parameter $\vec{v}$)
\begin{equation}
  K\left( \left\{ \vec{x},t\right\} \right) =K\left( \left\{ \vec{x}+2D\vec{v}
  t,t\right\} \right) e^{\vec{v}\cdot \left( \vec{x}_{1}+\vec{x}_{2}-
  \vec{x}_{3}-\vec{x}_{4}\right)} \ . \label{e}
\end{equation}
Clearly, a fully local interaction or the correlated noise term both 
proportional at least to $\delta \left( \vec{x}_{1}-\vec{x}_{3}\right)
\delta \left( \vec{x}_{2}-\vec{x}_{4}\right) $ fulfill Eq.~(\ref{e}). After
Fourier transformation, we get 
\begin{equation}
  \hat{K}\left( \left\{ \vec{p},\omega \right\} \right) =\hat{K}\left( \left\{ 
  \vec{p}-\dot{\imath}\vec{v},\omega +2D{\bf v\cdot p}\right\} \right) \ , 
\label{f}
\end{equation}
where, from translational invariance, $\vec{p}_{1}+\vec{p}_{2}=\vec{p}_{1}+
\vec{p}_{2}=:\vec{q}$ and $\omega _{1}+\omega _{2}=\omega _{1}+\omega_{2}=:
\omega $. Defining outgoing and ingoing wave vector transfers $\vec{k}$,
$\vec{k}^{\prime }$ as in Fig.~\ref{vertex}, accompanied with corresponding 
frequency transfers $\nu $ and $\nu ^{\prime }$, the function $\hat{K}$ can be 
reduced by Eq.~(\ref{e}) to the form
\begin{equation}
  \hat{K}\left( \left\{ \vec{p},\omega \right\} \right) =\hat{K}\left( \vec{k},
  \vec{k}^{\prime };\dot{\imath}\nu +D\vec{q}\vec{k},\dot{\imath}\nu ^{\prime}
  +D\vec{q} \vec{k}^{\prime },2\dot{\imath}\omega +D\vec{q}^{2}\right) \ .
\label{g}
\end{equation}
Causality shows that $\hat{K}$ is an analytic function of $\dot{\imath} \omega$
in the upper complex half-plane, whereas the analytic properties with respect 
to the variables $\nu $, $\nu^{\prime }$ are more involved. If $K\left(
\left\{ \vec{x},t\right\} \right) \propto \delta \left( t_{1}-t_{2}\right)
\delta \left( t_{3}-t_{4}\right) $, the Fourier transform of the function 
$\hat{K}$ is independent of the frequency transfers. Then Eq.~(\ref{f}) implies
\begin{equation}
  \hat{K}\left( \left\{ \vec{p},\omega \right\} \right) =\hat{K}\left( \vec{k},
  \vec{k}^{\prime };2\dot{\imath}\omega +D\vec{q}^{2}\right) \ .  \label{h}
\end{equation}
Because all terms contributing to the Bethe--Salpeter equation (\ref{betsal}) 
display such a time dependence, we find especially
\begin{eqnarray}
  &&\Gamma _{2,2}\left( \vec{k},\vec{k}^{\prime };\frac{2\dot{\imath}\omega }
  {D}+\vec{q}^{2}\right) \nonumber \\
  &&\quad = \hat{R}\left( \vec{k}-\vec{k}^{\prime }\right) 
  +\Gamma_{2,2}^{\left( D\right) }\left( \vec{k},\vec{k}^{\prime};
  \frac{2\dot{\imath} \omega }{D}+\vec{q}^{2}\right) \ , \label{i}
\end{eqnarray}
where $\Gamma_{2,2}^{\left( D\right) }$ is free of singularities in the
limit of vanishing arguments.



\begin{thebibliography}{10}
\bibitem{fnote2}
This is in complete analogy with the standard procedure in the static $\Phi^4$
  model, where the prefactor of the $(\nabla \Phi)^2$ term is conventionally
  set to $1$, and its renormalization is absorbed into a non-trivial field
  renormalization. The ensuing anomalous dimension eventually determines the
  static critical exponent $\eta$.
\bibitem{fnote3}
The exponent $\rho$ of the noise correlations must, however, must not be too
  large. For $\rho > (d+1)/2$, one enters a regime where $z$ would become less
  than $1$. In one dimension, Hayot and Jayaprakash
  \cite{hayot-jayaprakash:96,hayot-jayaprakash:97} discuss multi-fractal
  behavior and intermittency in that regime.
\end{thebibliography}
\end{document}